# Energy-Efficient Scheduling of HPC Applications in Cloud Computing Environments

Saurabh Kumar Garg [a,*], Chee Shin Yeo [b], Arun Anandasivam [c], Rajkumar Buyya [a]

[a] *Cloud Computing and Distributed Systems Laboratory, Department of Computer Science and Software Engineering, The University of Melbourne, Australia.*

[b] *Institute of High Performance Computing (IHPC), Agency for Science, Technology and Research (A*STAR), Singapore.*

[c] *Institute of Information Systems and Management, Karlsruhe Institute of Technology, Germany.*

**Abstract**

The use of High Performance Computing (HPC) in commercial and consumer IT applications is becoming popular. They need the ability to gain rapid and scalable access to high-end computing capabilities. Cloud computing promises to deliver such a computing infrastructure using data centers so that HPC users can access applications and data from a Cloud anywhere in the world on demand and pay based on what they use. However, the growing demand drastically increases the energy consumption of data centers, which has become a critical issue. High energy consumption not only translates to high energy cost, which will reduce the profit margin of Cloud providers, but also high carbon emissions which is not environmentally sustainable. Hence, energy-efficient solutions are required that can address the high increase in the energy consumption from the perspective of not only Cloud provider but also from the environment. To address this issue we propose near-optimal scheduling policies that exploits heterogeneity across multiple data centers for a Cloud provider. We consider a number of energy efficiency factors such as energy cost, carbon emission rate, workload, and CPU power efficiency which changes across different data center depending on their location, architectural design, and management system. Our carbon/energy based scheduling policies are able to achieve on average up to 30% of energy savings in comparison to profit based scheduling policies leading to higher profit and less carbon emissions.

## 1 Introduction

During the last few years, the use of High Performance Computing (HPC) infrastructure to run business and consumer based IT applications has increased rapidly. This is evident from the recent Top500 supercomputer applications where many supercomputers are now used for industrial HPC applications, such as 9.2% of them are used for Finance and 6.2% for Logistic services [1]. Thus, it is desirable for IT industries to have access to a flexible HPC infrastructure which is available on demand with minimum investment. Cloud computing [2] promises to deliver such reliable services through next-generation data centers [1] built on virtualized compute and storage technologies. Users are able to access applications and data from a "Cloud" anywhere in the world on demand and pay based on what they use. Hence, Cloud computing is a highly scalable and cost-effective infrastructure for running HPC applications which requires ever-increasing computational resources [3].

However, Clouds are essentially data centers which require high energy [2] usage to maintain operation [4]. Today, a typical data center with 1000 racks need 10 Megawatt of power to operate [5]. High energy usage is undesirable since it results in high energy cost. For a data center, the energy cost is a significant component of its operating and up-front costs [5]. Therefore, Cloud providers want to increase their profit or Return on Investment (ROI) by reducing their energy cost. Many Cloud providers are thus building different data centers and deploying them in many geographical location not only to expose their cloud services to business and consumer applications, e.g. Amazon [6] but also to reduce energy cost, e.g. Google [7].

In April 2007, Gartner estimated that the Information and Communication Technologies (ICT) industry generates about 2% of the total global $CO_2$ [3] emissions, which is equal to the aviation industry [8]. As governments impose carbon emissions limits on the ICT industry like in the automobile industry [9][10], Cloud providers must reduce energy usage to meet the permissible restrictions [11]. Thus, Cloud providers must ensure that data centers are utilized in a $CO_2$-efficient manner to meet scaling demand. Otherwise, building more data centers without any carbon consideration is not viable since it is not environmentally sustainable and will ultimately violate the imposed carbon emissions limits. This will in turn affect the future widespread adoption of

* Corresponding author.
*Email addresses:* sgarg@csse.unimelb.edu.au (Saurabh Kumar Garg), csyeo@ihpc.a-star.edu.sg (Chee Shin Yeo), anandasivam@kit.edu (Arun Anandasivam), raj@csse.unimelb.edu.au (Rajkumar Buyya).
[1] Data center and cloud site is used interchangeably.
[2] Energy and electricity is used interchangeably.
[3] $CO_2$ and carbon is used interchangeably.

Cloud computing, especially for the HPC community which demands scalable infrastructure to be delivered by Cloud providers. Companies like Alpiron[12] already offer software for cost-efficient server management and promise to reduce energy cost by analyzing, via advanced algorithms, which server to shutdown or turn on during the runtime.

Motivated by this practice, this paper enhances the idea of cost-effective management by taking both the aspects of economic (profit) and environmental (carbon emissions) sustainability into account. In particular, we aim to examine how a Cloud provider can achieve optimal energy sustainability of running HPC workloads across its entire Cloud infrastructure by harnessing the heterogeneity of multiple data centers geographically distributed in different locations worldwide.

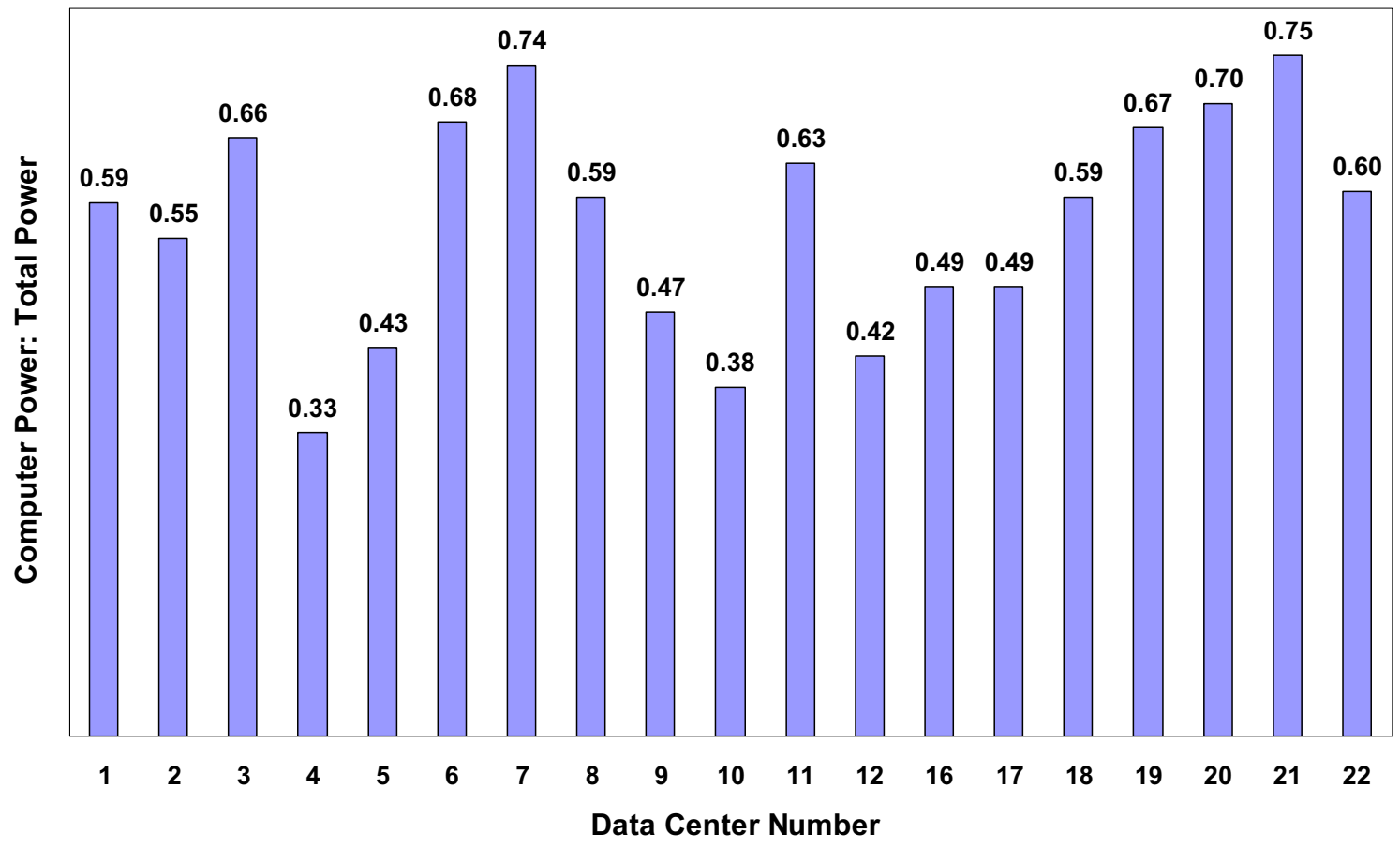

Fig. 1. Computer Power Consumption Index (Source: [13])

The analysis of previous work shows that little investigation has been done for both economic and environmental sustainability to achieve energy efficiency on a global scale as in Cloud computing. First, previous work has generally studied how to reduce energy usage from perspective of reducing cost, but not how to improve the profit while reducing the carbon emissions which is also significantly impacting the cloud providers [14]. Second, most previous work has focused on achieving energy efficiency at a *single* data center location, but not across *multiple* data center locations. However, Cloud providers such as Amazon EC2 [15] typically has multiple data centers distributed worldwide. As shown in Figure 1, the energy efficiency of an individual data center in different locations changes dynamically at various times depending on the number of factors such as energy cost, carbon emission rate, workload, CPU power efficiency, cooling system, and environmental temperature. Thus, these different contributing factors can be considered to exploit the heterogeneity across multiple data centers for improving the overall energy efficiency of the Cloud provider. Third, previous work has mainly proposed energy-saving policies that are application-specific [16][17], processor-specific [18][19], and/or server-specific [20][21]. But, these policies are only applicable or most effective for the specific models that they are specially designed for. Hence, we propose some simple, yet effective generic energy-efficient scheduling policies that can be extended to any other application model, processor and server models so that they can be readily deployed in existing data centers with minimum changes. Our generic scheduling policies within a data center can also easily complement any of these application-specific, processor-specific, and/or server-specific energy-saving policies that are already in place within existing data centers or servers.

The key contributions of this paper are:

(1) A novel mathematical model for energy efficiency based on various contributing factors such as energy cost, $CO_2$ emission rate, HPC workload, and CPU power efficiency.
(2) The near-optimal energy-efficient scheduling policies which not only minimizes the $CO_2$ emissions and maximizes the profit of the Cloud provider, but also can be readily implemented without much infrastructure changes such as the relocation of existing data centers.
(3) Energy efficiency analysis of our proposed policies (in terms of $CO_2$ emissions and profit) through extensive simulations using real HPC workload traces and data center $CO_2$ emission rates and energy costs to demonstrate the importance of considering various contributing factors.
(4) The analysis of lower/upper bounds of the optimization problem
(5) A novel Dynamic Voltage Scaling (DVS) based approach for scheduling HPC applications within a data center which can reduce energy consumption by about 30%.

The paper is organized as follows. Wection 2 the related work focusing on energy efficiency or cost- and market-based schedulers are briefly described. Section 3 defines the Cloud Computing scenario and the problem description. In Section 4 different kind of policies for allocating efficiently applications on machines are scrutinized. Section 5 explains the evaluation methodology and simulation setup for the energy sustainability analysis of various policies followed by the analysis of the results in Section 6. Section 7 presents the conclusion and future work.

## 2 Related Work

Many research work address energy-efficient computing for servers [4]. The most relevant to the context of Cloud computing are the following:

- **Cluster Servers**: Bradley et al. [22] proposed algorithms to minimize the power utilization by using workload history and predicting future workload within acceptable reliability. Chen et al. [23] simulated a cluster using both predictive queuing models and feedback controllers to derive frequency adjustments for web serving clusters over control intervals lasting several minutes. Lawson and Smirni [24] proposed an energy saving scheme that dynamically adjusts the number of CPUs in a cluster to operate in "sleep" mode when the utilization is low. Kim et al. [21] proposed power-aware scheduling algorithms for bag-of-tasks applications with deadline constraints on DVS-enabled cluster systems. Wang and Lu [20] presented a threshold-based approach for efficient power management of a single heterogeneous soft real-time cluster. Tesauro et al. [25] presented an application of batch reinforcement learning combined with nonlinear function approximation to optimize multiple aspects of data center behavior such as performance, power, and availability.

- **Virtualized Servers**: Nathuji and Schwan [26] integrated power management mechanisms and policies with virtualization technologies to reduce power consumption for web workloads. Verma et al. [27] proposed the placement of applications on heterogeneous virtualized servers based on power and migration cost. Cardosa et al. [28] exploited the minimum-maximum resource partitions and shares features in Virtual Machine (VM) technology to manage power in a data center.

These solutions target to save energy within a single server or a single data center (with many servers) in a *single* location. Since our generic scheduling policy improves the energy efficiency across data centers in *multiple* locations with different carbon emission rates, it can be used in conjunction with these solutions to utilize any energy efficiency already implemented in a single location.

There are some studies on energy efficiency in Grids, which comprise resource sites in multiple locations similar to our scope. Orgerie et al. [29] proposed a prediction algorithm to reduce power consumption in large-scale computational grids such as Grid5000 by aggregating the workload and turning off unused CPUs. Hence, they do not consider using DVS to save power for CPUs. Patel et al. [30] proposed allocating Grid workload on a global scale based on the energy efficiency at different data centers. But, their focus is on reducing temperature, and thus do not examine how energy consumption can be reduced by exploiting different power efficiency of CPUs, energy costs, and carbon emission rates across data centers. In addition, they do not focus on any particular workload characteristics, whereas we focus on HPC workload.

Most previous work focuses on reducing energy consumption in data centers for web workloads [20][23]. Thus, they assume that energy is an increasing function of CPU frequency since web workloads have the same execution time per request. However, HPC workloads have different execution time depending on specific application requirements. Hence, the energy-CPU frequency relationship of a HPC workload is significantly different from that of a web workload as discussed in Section 4.2. Therefore, in this paper, we define a generalized power model and adopt a more general strategy to scale up or down the CPU frequency.

Not many research work studies the energy sustainability issue from an economic cost perspective. To address energy usage, Chase et al. [31] adopted an economic approach to manage shared server resources in which services "bid" for resources as a function of delivered performance. Burge et al. [32] scheduled tasks to heterogeneous machines and made admission decisions based on the energy costs of each machine to maximize the profit in a single data center. None of these research work studies the critical relationship between $CO_2$ emissions (environmental sustainability) and profit (economic sustainability) for the energy sustainability issue, and how they can affect each other. On the other hand, we examine how the carbon emissions can be reduced for executing HPC applications with negligible effect on the profit of the Cloud provider.

Market-oriented meta-schedulers [33] have been proposed to consider cost constraints. Kumar et al. [34] optimizes the assignment of applications by maximizing the business value with respect to the time and cost of users. Garg et al. [35] examines how parallel applications can be executed most economically in the minimum time by managing the trade-off between time and cost constraints. However, none of these meta-schedulers aims to reduce $CO_2$ emissions for environmental sustainability, in addition to meeting cost objectives.

## 3 Meta-scheduling Model

### *3.1 System Model*

Our system model is based on the Cloud computing environment, whereby Cloud users are able to tap the computational power offered by the Cloud providers to execute their HPC applications. The Cloud meta-scheduler acts as an interface to the Cloud infrastructure and schedules applications on the behalf of users as shown in Figure 2. It interprets and analyzes the service requirements of a submitted application and decides whether to accept or reject the application based on the availability of CPUs. Its objective is to schedule applications such that the $CO_2$ emissions can be reduced and the profit can be increased, while the Quality of Service (QoS) requirements of the applications are met. As data centers are located in different geographical regions, they have different $CO_2$ emission rates and energy costs depending on regional constraints. Each data center is responsible for updating this information to the meta-scheduler for energy-efficient scheduling. The two participating parties, Cloud users and Cloud providers, are discussed below along with their objectives and constraints:

(1) **Cloud Users:** The Cloud users need to run HPC applications/workloads which is compute-intensive and thus require parallel and distributed processing to significantly reduce the processing time. The users submit parallel applications with QoS requirements to the Cloud meta-scheduler. Each application must be executed in an individual data center and does not have preemptive priority. The reason for this requirement is that the synchronization among various tasks of parallel applications can be affected by communication delays when applications are executed across multiple data centers. The user's objective is to have their application completed by a deadline. Deadlines are hard, i.e., the user will benefit from the HPC resources only if the application completes before its deadline [36]. To facilitate the comparison between the policies described in this work, the estimated execution time of an application provided by the user is considered to be accurate [37]. Several models, such as those proposed by Sanjay and Vadhiyar [38], can be applied to estimate the runtime of parallel applications. In this work, an application execution time is inversely proportional to the CPU operating frequency.

(2) **Cloud Providers:** A Cloud provider has multiple data centers distributed across the world. For example, Amazon [6] has data centers in many cities across Asia, Europe, and United States. Each data center has a local scheduler that manages the execution of incoming applications. The Cloud meta-scheduler interacts with these local schedulers for application execution. Each local scheduler periodically supplies information about available time slots $(t_s, t_e, n)$ to the meta-scheduler, where $t_s$ and $t_e$ are the start time and end time of the slot respectively and $n$ is the number of CPUs available for the slot. To facilitate energy-efficient computing, each local scheduler also supplies information about the $CO_2$ emission rate, Coefficient of Performance (COP), CPU power-frequency relationship, Million Instructions Per Second (MIPS) rating of CPUs at the maximum frequency, and CPU operating frequency range of the data center. The MIPS rating is used to indicate the overall performance of CPU. All CPUs within a data center are homogeneous, but CPUs can be heterogeneous across data centers.

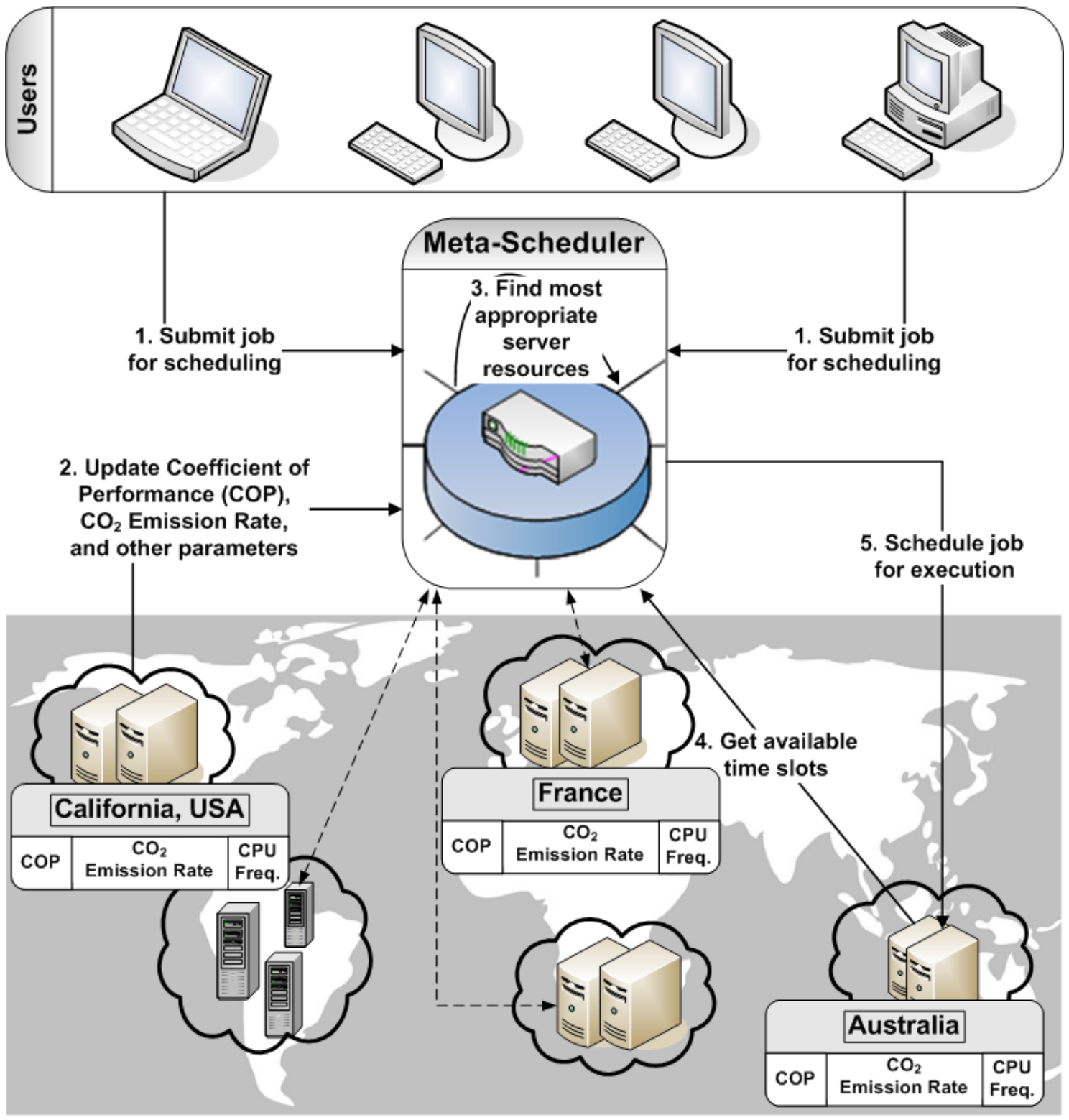

Fig. 2. Cloud meta-scheduling protocol

### *3.2 Data Center Energy Model*

The major contributors for the total energy usage in a data center are computing devices (CPUs) and cooling system which constitute about 80% of the total energy consumption. Other systems such as lighting are not considered due to their negligible contribution to the total energy cost [39].

Power consumption can be reduced by lowering the supply voltage of CPUs using DVS. DVS is an efficient way to manage dynamic power dissipation during computation. The power consumption $P_i$ of a CPU in a data center at cloud site $i$ is composed of dynamic and static power [23][20]. The static power includes the base power consumption of the CPU and the power consumption of all other components. Thus, the CPU power $P_i$ is approximated by the following function (similar to previous work [23][20]): $P_i = \beta_i + \alpha_i f^3$, where $\beta_i$ is the static power consumed by the CPU, $\alpha_i$ is the proportionality constant, and $f$ is the frequency at which the CPU is operating. We consider that a CPU of a data center can adjust its frequency from a minimum of $f_{i,min}$ to a maximum of $f_{i,max}$ discretely.

The energy cost of the cooling system depends on the Coefficient Of Performance (COP) factor [40]. COP is an indication for the efficiency of cooling system, which is defined as the ratio of the amount of energy consumed by CPUs to the energy consumed by the cooling system. The COP is, however, not constant and varies with cooling air temperature. We assume that COP will remain constant during scheduling cycle and resource sites will update meta-scheduler whenever COP changes. Thus, the total energy consumed by cooling system is given by:

$$E_{h,i} = \frac{E_{c,i}}{COP_i} \quad (1)$$

where $E_{c,i}$ is the total energy consumed in computing devices and $E_{h,i}$ is the total energy consumed by cooling devices.

### *3.3 Relation between Execution Time and CPU frequency*

Since, we have considered DVS mechanism to scale up and down CPU frequency, the execution time of user application will significantly vary according to the CPU frequency. The decrease in execution time due to increase in CPU frequency depends on whether application is CPU bound or not. For example, if an application's performance is completely dependent on the CPU frequency, then its execution time will be inversely proportional to the change in CPU frequency. Thus, the execution time of an application is modeled according to the definition proposed by Hsu et al. [41]:

$$T(f) = T(f_{max}) \times (\gamma_{cpu}(\frac{f_{max}}{f} - 1) + 1) \quad (2)$$

where $T(f)$ is the execution time of application at CPU frequency $f$, and $T(f_{max})$ is the execution time of running at the top frequency $f_{max}$.

The parameter $\gamma_{cpu}$ reflects CPU boundness of an application. If the value of $\gamma_{cpu}$ decreases, the CPU-boundness of the application will also decrease, which results in potentially more energy reduction by using DVS. The worst case value for $\gamma_{cpu}$, i.e. $\gamma_{cpu} = 1$, is assumed to analyze the performance of our heuristics.

### *3.4 Problem Description*

Let a Cloud provider have $N$ data centers distributed in different sites. All the parameters associated with a data center $i$ are given in table 1. A user submit his/her requirements for an application $j$ in the form of a tuple $(d_j, n_j, e_{j1}.., e_{jN}, \gamma_{cpu,j})$, where $d_j$ is the deadline to complete application $j$, and $n_j$ is the number of CPUs required for application execution, $e_{ji}$ is the application execution time on the data center $i$ when operating at the maximum CPU frequency, and $\gamma_{cpu,j}$ is CPU boundness. In addition, let $f_{ij}$ be the initial frequency at which CPUs of a data center $i$ operate while executing application $j$. Executing application $j$ on data center $i$ will result in the following:

(1) Energy consumption by CPU

$$E^c_{ij} = (\beta_i + \alpha_i f^3_{ij}) \times n_j e_{ji} \times (\gamma_{cpu,j}(\frac{f_{i,max}}{f_{ij}} - 1) + 1) \tag{3}$$

(2) Total energy which consist of cooling system and CPU

$$E_{ij} = (1 + \frac{1}{COP_i})E^c_{ij} \tag{4}$$

(3) Energy cost

$$C^e_{ij} = E_{ij} \times c^e_i \tag{5}$$

(4) $CO_2$ emission

$$(CO2E)_{ij} = r_i^{CO_2} \times E_{ij} \tag{6}$$

(5) Profit

$$(Prof)_{ij} = e_{ji} n_j p_i - C^e_{ij} \tag{7}$$

Thus the meta-scheduling problem can be formulated as

$$\text{Minimize CO2E} = \sum_i^N \sum_j^J x_{ij} r_i^{CO_2} E_{ij} \tag{8}$$

$$\text{Maximize Profit} = \sum_i^N \sum_j^J (Prof)_{ij} x_{ij} \tag{9}$$

Subject to:

(1) Response time of application $j < d_j$
(2) $f_i^{min} < f_{ij} < f_i^{max}$
(3) $\sum_i^N x_{ij} \leq 1$
(4) $$x_{ij} = \begin{cases} 1 & \text{if application } j \text{ allocated to data center } i \\ 0 & \text{otherwise} \end{cases}$$

The dual objective functions (8) and (9) of the meta-scheduling problem are to minimize the $CO_2$ emission and maximize revenue of a resource provider. The constraint (1) forces to meet the deadline requirement of an application. It is difficult to calculate the exact response time of an application in the environment where applications have different sizes and require multiple CPUs and has very dynamical arrival rate [23]. Moreover, this problem maps to 2-dimensional bin packing problem which is NP-hard in nature [42]. Thus, we propose various scheduling policies to heuristically approximate the optimum.

## 4 Meta-Scheduling Policies

The meta-scheduler periodically assigns applications to cloud sites. The time period between two scheduling cycles is called the *scheduling cycle*. In each scheduling cycle, the meta-scheduler collects the information from both cloud sites and users. All parameters associated with each resource site $i$ are given in Table 1.

Table 1
**Parameters of a Cloud Site $i$**

| Parameter | Notation |
|---|---|
| $CO_2$ emission rate (kg/kWh) | $r_i^{co2}$ |
| Average COP | $COP_i$ |
| Electricity cost ($/kWh) | $c_i^e$ |
| Execution Price ($/sec) | $p_i$ |
| CPU power | $P_i = \beta_i + \alpha_i f^3$ |
| CPU frequency range | $[f_{i,min}, f_{i,max}]$ |
| Time slots (start time, end time, number of CPUs) | $(t_s, t_e, n)$ |

In general, a meta-scheduling policy consists of two phases: 1) mapping phase, in which the meta-scheduler first maps an application to a cloud site; and 2) scheduling phase, in which the scheduling of applications is done within the data center of selected cloud site, where the required time slots is chosen to complete the application. Depending on the objective of cloud provider, whether he want to minimize carbon emission or maximize profit, we have designed various mapping policies which are discussed in the subsequent section. To further reduce the energy consumption within the data center, we have designed a DVS based scheduling for local scheduler of the data center at each cloud site.

### 4.1 Polices for Mapping Jobs to Resource

We have designed the following meta-scheduling mapping/allocation policies depending on the objective of the cloud provider:

#### 4.1.1 Minimizing Carbon Emission

The following policies optimize the global carbon emission by all cloud sites while keeping number of missed deadlines low.

- **Greedy Minimum Carbon Emission (GMCE):** In this greedy based policy, all user applications are ordered by their deadline (earliest first), while the data centers at different cloud sites are sorted in decreasing order of their Carbon efficiency i.e. $r_i^{CO_2} \times ((\beta_i/f_{i,max}) + \alpha_i(f_{i,max})^2)(\frac{1+COP_i}{COP_i})$. Then, the meta-scheduler assigns applications to a cloud site according to this ordering.
- **Minimum Carbon Emission Minimum Carbon Emission (MCE-MCE):** MCE-MCE is based on the general concept of the Min-Min idea [43]. The Min-Min type heuristic performed very well in previous studies of different environments (e.g., [44]). In, MCE-MCE, the meta-scheduler finds the best data center or cloud sites for all applications that are considered, and then among these applications/sites pairs, the meta-scheduler selects the best pair to map first. To determine which application/cloud site pair is the best, we have used the carbon emission resulted due to execution of the application $j$ on the cloud site $i$ i.e. $(CO2E)_{ij}$ as the fitness value. Thus, MCE-MCE includes the following steps:

  **Step 1:** For each application that is present in the meta-scheduler to schedule, find the cloud site for which carbon emission is the minimum, i.e. minimum fitness value (the first MCE) among all cloud sites which can complete the application by its deadline. If there is no cloud site where the application can be completed by its deadline, then the application is dropped or removed from the list of applications which are to be mapped.

  **Step 2:** Among all the application/cloud site pairs found in Step 1, find the pair that resulted in minimum carbon emission, i.e. minimum fitness value (the second MCE). Then, map the application to the resource site, and remove the application from the list of applications which are to be mapped.

  **Step 3:** Update the available slots from the resource sites.

  **Step 4:** Do steps 1 to 3 until all applications are mapped.

### *4.1.2 Maximizing Profit*

Following policies optimize the global profit gained by cloud provider while keeping the number of deadlines miss low.

- **Greedy Maximum Profit (GMP):** This policy maps the incoming applications to most profitable cloud site for maximum profit. Thus, all user applications are ordered by their deadline (earliest first), while cloud sites are ordered by their electricity cost i.e. $c_i^e \times ((\beta_i/f_{i,max}) + \alpha_i(f_{i,max})^2)(\frac{1+COP_i}{COP_i})$. Then, meta-scheduler assigns applications to cloud sites according to this ordering.
- **Maximum Profit Maximum Profit(MP-MP):** MP-MP is very similar to the general concept of the Min-Min idea [43]. In, MP-MP, the meta-scheduler finds the cloud sites for all applications that are considered which results in the maximum profit, and then among these applications/sites pairs, the meta-scheduler selects the pair to map first, which is the most profitable. To determine which application/cloud site pair is the most profitable, we have used the profit resulted due to execution of the application $j$ on the cloud site $i$, i.e. $(Prof)_{ij}$ as the fitness value. Thus, MP-MP includes the following steps:

  **Step 1:** For each application that is present in the meta-scheduler for execution, find the cloud site for which profit is the maximum i.e. maximum fitness value (the first MP) among all the cloud sites which can complete the application by its deadline. If there is no resource site where the application can be completed by its deadline, then the application is dropped or removed from the list of applications which are to be mapped.

  **Step 2:** Among all the application/cloud site pairs found in Step 1, find the pair that resulted in maximum profit, i.e. maximum fitness value (the second MP). Then, map the application to the cloud site, and remove the application from the list of applications which are to be mapped.

  **Step 3:** Update the available slots from the cloud sites.

  **Step 4:** Do steps 1 to 3 until all applications are mapped.

### *4.1.3 Minimizing Carbon Emission and Maximizing Profit (MCE-MP)*

In this policy, the objective of the meta-scheduler is to minimize the total carbon emission while maximizing the total profit of the Cloud provider. Thus, this policy handles the trade-off between profit and carbon emission which may be conflicting. This policy is very similar to MCE-MCE and MP-MP except fitness functions (carbon emission i.e. $(CO2E)_{ij}$ and profit $(Prof)_{ij}$) for each step of finding "best" application/cloud site pair. Thus, the MCE-MP policy include following steps:

**Step 1:** For each of the applications that are present in meta-scheduler for execution, find the cloud site for which the carbon emission is minimum, i.e. minimum $(CO2E)_{ij}$ (the first MCE) among all the cloud sites that can complete the application by its deadline. If there is no such cloud site where the application can be completed by its deadline, then the application is removed from the list of applications that are to be mapped.

**Step 2:** Among all the application/cloud site pairs found in Step 1, find the pair that resulted in maximum profit i.e. maximum $(Prof)_{ij}$ (the second MP). Then, map the application to the cloud site and remove the application from the list of applications that are to be mapped.

**Step 3:** Update the available slots from the cloud sites.

**Step 4:** Do steps 1 to 3 until all applications are mapped.

### *4.2 Scheduling Policy*

The energy consumption and carbon emission are further reduced within a data center by using DVS at the CPU level that can save energy by scaling down the CPU frequency. Thus, before the meta-scheduler assigns an application to a cloud site, it decides the time slot in which the application should be executed and the frequency at which the CPU should operate to save energy. But, since a lower CPU operating frequency can increase the number of application rejected due to the deadline misses, the scheduling of the HPC applications within the data center can be of two types: 1) CPUs run at the maximum frequency (i.e. without DVS) or 2) CPUs run at various frequency using DVS (i.e. with DVS). It is important to adjust DVS appropriately in order to reduce the number of missed deadlines and energy consumption simultaneously.

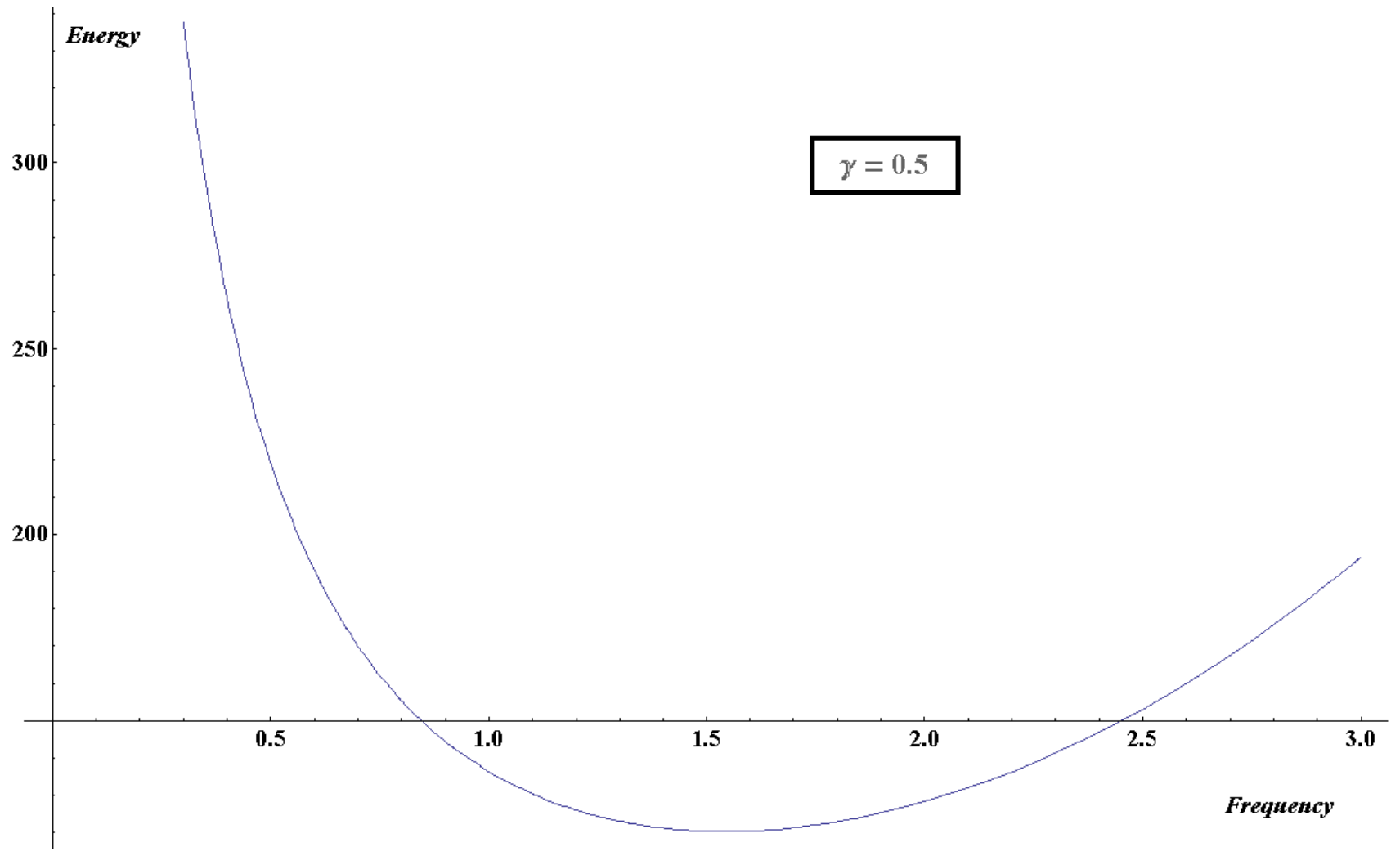

Fig. 3. Energy Consumption VS Frequency

Section 3.4 shows that the energy consumption by a CPU depends on fre-quency at which an application will be executed. The objective is to obtain an optimal frequency so that the energy consumption can be minimized while executing the application within its deadline. From the plot of energy consumption in Figure 3, we can observe existence of local minima where energy consumption will be minimum. In order to identify this local minimum we differentiate energy consumption of a HPC application $j$ on a CPU at a data center $j$ with respect to operating CPU frequency $f_{ij}$ ,

$$E^c_{ij} = (\beta_i + \alpha_i f^3_{ij}) \times n_j e_{ji} \times (\gamma_{cpu,j}(\frac{f_{i,max}}{f_{ij}} - 1) + 1) \tag{10}$$

$$\frac{\partial(E^c_{ij})}{\partial f_{ij}} = n_j e_{ji} \times (\frac{((\beta_i + \alpha_i f^3_{ij}) f_{i,max} \gamma_{cpu,j})}{f^2_{ij}} + 3\alpha_i f^2_{ij}(1 + (-1 + \frac{f_{i,max}}{f_{ij}})\gamma_{cpu,j})) \tag{11}$$

For local minima,

$$\frac{\partial(E^c_{ij})}{\partial f_{ij}} = 0 \tag{12}$$

$$\frac{((\beta_j + \alpha_i f^3_{ij}) f_{i,max} \gamma_{cpu,j})}{f^2_{ij}} + 3\alpha_i f^2_{ij}(1 + (-1 + \frac{f_{i,max}}{f_{ij}})\gamma_j) = 0 \tag{13}$$

In Figure 3, we can clearly see that local minima exist. Thus, at least one root of above polynomial will exist in range $[0, \infty]$. When $\gamma_{cpu,j} = 1$, the above equation will reduce to,

$$(\frac{-\beta_i}{f^2_{ij}} + 2\alpha_i f_{ij}) = 0 \tag{14}$$

It can be noted that the above equation depends on static variables such as CPU power efficiency. Thus, we can pre-compute the local minima before starting meta-scheduling algorithm.

The resulting optimal CPU frequency $f_{i,opt}$ is not bounded to $[f_{i,min}, f_{i,max}]$. Since the CPU frequency of resource site $i$ can operate only in the interval $[f_{i,min}, f_{i,max}]$, we define $f_{i,opt} = f_{i,min}$, if $f_{i,opt} < f_{i,min}$, and $f_{i,opt} = f_{i,max}$, if $f_{i,opt} > f_{i,max}$.

The meta-scheduler will first try to operate the CPU at a frequency in the range $[f_{i,min}, f_{i,max}]$ nearest to $f_{i,opt} = \sqrt[3]{\frac{\beta_i}{2\alpha_i}}$. If the deadline of an application will be violated, the meta-scheduler will scale up the CPU frequency to the next level and then again try to find the free slot to execute the application. If the meta-scheduler fails to schedule the application on the cloud site as no free slot is available, then the application is forwarded to the next cloud site for scheduling (the ordering of resource sites is described in Section 4.1).

### *4.3 Upper Bound and Lower Bound*

Due to the NP hardness of the meta-scheduling problem described in Section 3.4, it is difficult to find the optimal profit and the carbon emission in polynomial time. To estimate the performance of our scheduling algorithms we present an upper bound of cloud provider profit and a lower bound of carbon emissions due to execution of HPC workload for comparison purpose. The bounds are based on the principle that we can get the maximum profit or the minimum carbon emission when most of the applications run on the most "efficient" data center and also at the optimal frequency. The most "efficient" data center for the profit maximization is the data center that results in the minimum energy cost for executing HPC applications. Similarly, the most "efficient" data center for the carbon emission minimization is the data center that causes the minimum carbon emission for executing HPC applications.

To map the maximal number of HPC applications to the most efficient data center, we relaxed some constraints of our system model. First, we relaxed the constraint that when an application is executed at highest CPU frequency it will result in the maximum energy consumption. To calculate the lower and upper bounds we assume that all applications will be executed at the highest CPU frequency, while for computing the energy consumption due to application we will take into account energy consumed at the optimal CPU frequency. Second, even though HPC applications considered in the system model are parallel applications with fixed CPU requirements, we relaxed this constraint to applications that are moldable in required number of CPUs. Thus, the runtime of application will decrease linearly when it is scheduled on a bigger number of CPUs. This will increase the number of applications that can be allocated to the most efficient site with the minimum energy possible. Third, the applications in the system are arriving dynamically in different scheduling cycles, but for the identification of the bounds all applications are considered in one scheduling cycle and then mapped them to data centers while considering their deadlines. This is the best case, when all information about the incoming applications are known in advance. Of course, the dynamic system will have a worse performance.

The bounds of carbon emission and profit, thus, obtained by having these assumption are unreachable loose bounds of the system model considered since with these assumptions, data center will be executing the maximum possible workload with 100% utilization of CPUs while the least possible energy consumption is considered for comparison. Let TWL be the total workload scheduled, TP be the total profit gained by cloud provider, and TCE be the total carbon emission. The computation of lower bounds involve the following steps:

**Step 1:** All HPC applications are ordered by their deadline, while the data centers in all cloud sites are sorted in decreasing order of their carbon emission efficiency, i.e. $r_i^{CO_2} \times ((\beta_i/f_i^{max})+\alpha_i(f_i^{max})^2)(\frac{1+COP_i}{COP_i})$. Each application will be mapped to a data center in this ordering.

**Step 2:** For each HPC application $j$, search for the data center $i$, starting from the most efficient one, where the application $j$ can be scheduled without missing deadline when running at maximum CPU frequency.

**Step 3:** If data center $i$ is not found, then the application $i$ will be removed from the potential application list. Go to step 2 for scheduling other applications.

**Step 4:** If data center $j$ is found, application $i$ is scheduled and molded such that there is no fragmentation in the data center schedule for executing applications.

**Step 5:** $TWL+ = n_j * e_j i$

**Step 5:** $TCE+ = r_i^{CO_2} n_j$(Execution time of application $j$ at optimal frequency)$\times$(Power consumption at optimal CPU frequency)$(\frac{1+COP_i}{COP_i})$.

**Step 7:** $TP+ = n_j*$(Execution time of application $j$ at optimal frequency)*(1-$c_i^e\times$(Power consumption at optimal CPU frequency)$(\frac{1+COP_i}{COP_i})$)

**Step 8:** Repeat from Step 2 unless all applications are scheduled.

$\frac{TCE}{TWL}$ will be the lower bound of the average carbon emission due to the execution of all HPC application across multiple data center of cloud provider. For finding the upper bound of the average profit gained by cloud provider, in the first step, the sorting of data center is done in increasing order of the energy cost efficiency i.e., $c_i^e \times ((\beta_i/f_i^{max}) + \alpha_i(f_i^{max})^2)(\frac{1+COP_i}{COP_i})$ instead of carbon emission efficiency, otherwise other steps will remain same.

## 5 Performance Evaluation

We use workload traces Feitelson's Parallel Workload Archive (PWA) [45] to model the HPC workload. Since this paper focuses on studying the application requirements of cloud users with HPC applications, the PWA meets our objective by providing application traces that reflect the characteristics of real parallel applications. The experiments utilize the applications in the first week of the LLNL Thunder trace (January 2007 to June 2007). The LLNL Thunder trace from the Lawrence Livermore National Laboratory (LLNL) in USA is chosen due to its highest resource utilization of 87.6% among available traces to ideally model a heavy workload scenario. From this trace, we obtain the submit time, requested number of CPUs, and actual runtime of applications. However, the trace does not contain the service requirement of applications (i.e. deadline). Hence, we use a methodology proposed by Irwin et al. [46] to synthetically assign deadlines through two classes namely Low Urgency (LU) and High Urgency (HU).

An application $i$ in the LU class has a high ratio of $deadline_i/runtime_i$ so that its deadline is definitely longer than its required runtime. Conversely, an application $i$ in the HU class has a deadline of low ratio. Values are normally distributed within each of the high and low deadline parameters. The ratio of the deadline parameter's high-value mean and low-value mean is thus known as the high:low ratio. In our experiments, the deadline high:low ratio is 3, while the low-value deadline mean and variance is 4 and 2 respectively. In other words, LU applications have a high-value deadline mean of 12, which is 3 times longer than HU applications with a low-value deadline mean of 4. The arrival sequence of applications from the HU and LU classes is randomly distributed.

Table 2
**Characteristics of Cloud Sites**

| Location | $CO_2$ Emission Rate [b] (kg/kWh) | Energy Cost [a] (\$/kWh) | CPU Power Factors | | CPU Frequency Level | | Number of CPUs |
|---|---|---|---|---|---|---|---|
| | | | $\beta$ | $\alpha$ | $f_i^{max}$ | $f_i^{opt}$ | |
| New York, USA | 0.389 | 0.15 | 65 | 7.5 | 1.8 | 1.630324 | 2050 |
| Pennsylvania, USA | 0.574 | 0.09 | 75 | 5 | 1.8 | 1.8 | 2600 |
| California, USA | 0.275 | 0.13 | 60 | 60 | 2.4 | 0.793701 | 650 |
| Ohio, USA | 0.817 | 0.09 | 75 | 5.2 | 2.4 | 1.93201 | 540 |
| North Carolina, USA | 0.563 | 0.07 | 90 | 4.5 | 3.0 | 2.154435 | 600 |
| Texas, USA | 0.664 | 0.1 | 105 | 6.5 | 3.0 | 2.00639 | 350 |
| France | 0.083 | 0.17 | 90 | 4.0 | 3.2 | 2.240702 | 200 |
| Australia | 0.924 | 0.11 | 105 | 4.4 | 3.2 | 2.285084 | 250 |

[a] Energy cost reflects average commercial rates till 2007 based on a US Energy Information Administration (EIA) report [47].
[b] $CO_2$ emission rates are derived from a US Department of Energy (DOE) document [48] (Appendix F-Electricity Emission Factors 2007).

**Provider Configuration:** We model 8 different cloud sites (data centers) with different configurations as listed in Table 2. Power parameters (i.e. CPU power factors and frequency level) of the CPUs at different sites are derived from Wang and Lu's work [20]. Current commercial CPUs only support discrete frequency levels, such as the Intel Pentium M 1.6GHz CPU, which supports 6 voltage levels. We consider discrete CPU frequencies with 5 levels in the range $[f_{i,min}, f_{i,max}]$. For the lowest frequency $f_{i,min}$, we use the same value used by Wang and Lu [20], i.e. $f_{i,min}$ is 37.5% of $f_{i,max}$. Each local scheduler at a cloud site use Conservative Backfilling with advance reservation support as used by Mu'alem and Feitelson [49]. The meta-scheduler schedules the application periodically at a scheduling interval of 50 seconds, which is to ensure that the meta-scheduler can receive at least one application in every scheduling interval. The COP (power usage efficiency) value of resource sites is randomly generated using a uniform distribution between [0.6, 3.5] as indicated in the study conducted by Greenberg et al. [13]. The cloud provider's price to execute an application is fixed at 40 cents/hour.

**Performance Metrics:** Four metrics are necessary to compare the policies:

average energy consumption, profit gained, workload executed and average $CO_2$ emission. The *Average energy consumption* compares the amount of energy saved by using different scheduling algorithms, whereas the *average $CO_2$ emission* compares its corresponding environmental impact. Since minimizing carbon emission can effect cloud provider economically by decreasing his/her profit, we have considered *profit gained* as another metric to compare different algorithms. It is important to know the effect of various meta-scheduling policies on energy consumption, since higher energy consumption is likely to generate more $CO_2$ emission for worse environmental impact and incur more energy cost for operating cloud sites.

**Experimental Scenarios:** We examine various experimental scenarios to evaluate the performance of our algorithms:

- Effect of urgency and arrival rate of applications
- Effect of mapping policy and DVS
- Impact of carbon emission rate
- Impact of electricity cost
- Impact of data center efficiency
- Comparison with lower bound and upper bound
- Comparison of our DVS and previous DVS

For the above scenarios, we observe the performance from the user and provider perspective. From the user perspective, we observe the performance of varying: 1) urgency class and 2) arrival rate of applications. For the urgency class, we use various percentages (0%, 20%, 40%, 60%, 80%, and 100%) of HU applications. For instance, if the percentage of HU applications is 20%, then the percentage of LU applications is the remaining 80%. For the arrival rate, we use various factors (10 (low), 100 (medium), 1000 (high), and 10000 (very high)) of submit time from the trace. For example, a factor of 10 means an application with a submit time of 10s from the trace now has a simulated submit time of 1s. Hence, a higher factor represents higher workload by shortening the submit time of applications.

## 6 Analysis of Results

### *6.1 Effect of Urgency and Arrival Rate of Applications*

Figure 4 shows how the urgency and arrival rate of applications affects the performance of carbon emission based policies (GMCE, MCE-MCE, and MCE-MP) and profit based policies (GMP and MP-MP). The metrics of total carbon emission and total profit are used since the Cloud provider needs to know the

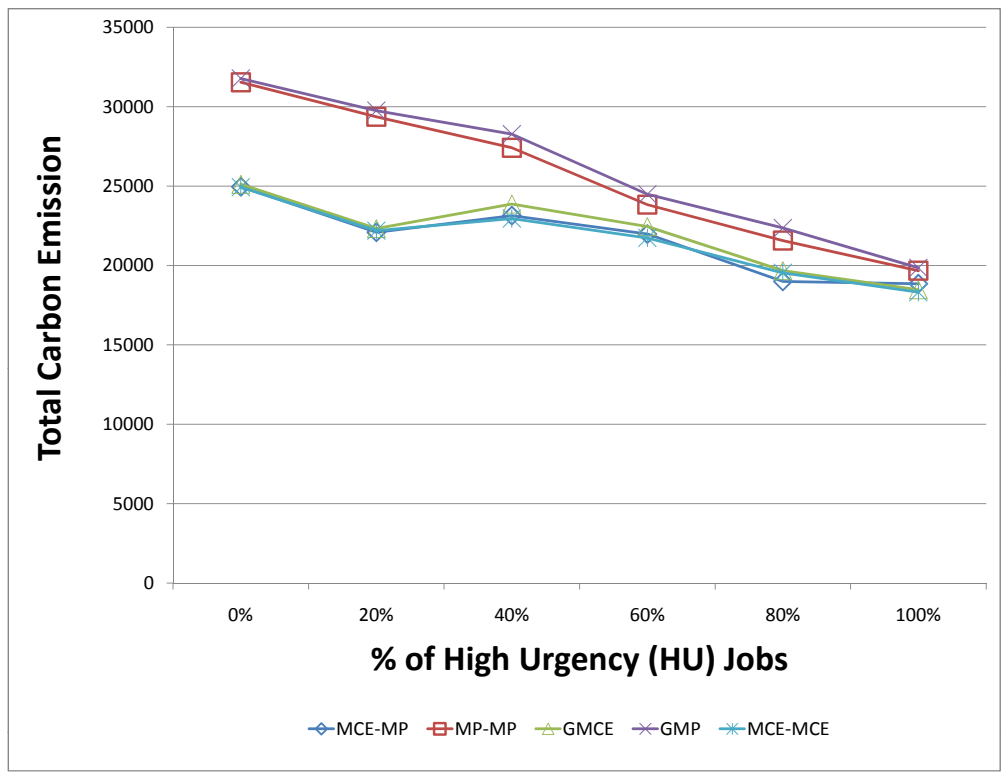

(a) Carbon Emission VS Urgency

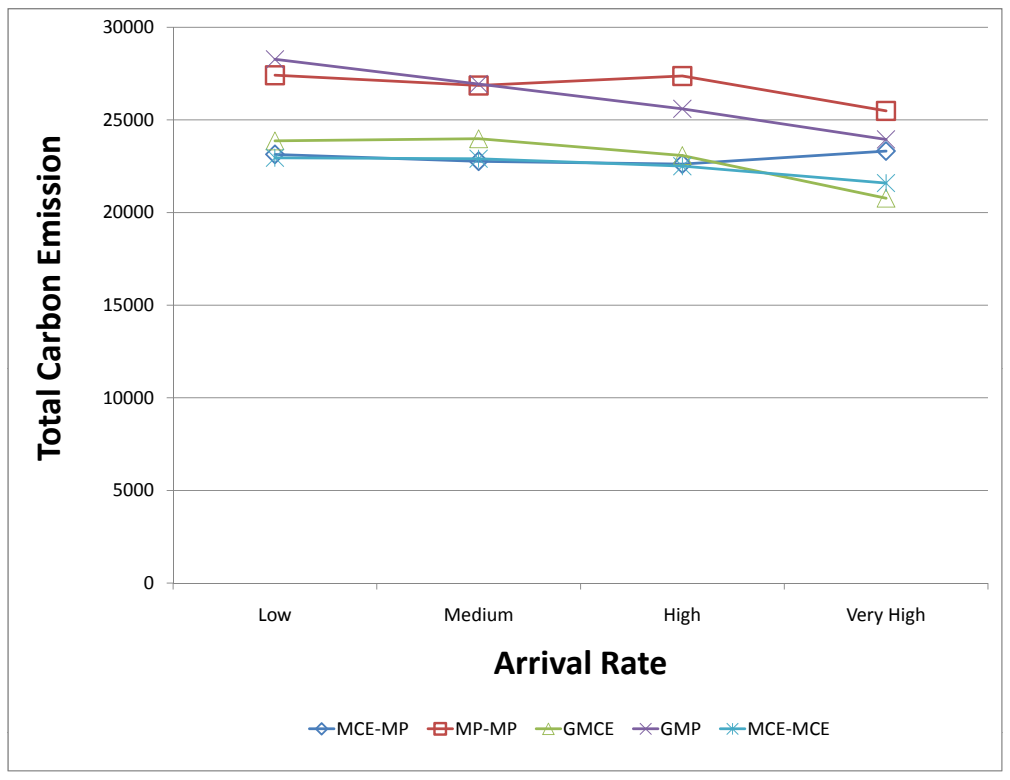

(b) Carbon Emission VS Arrival Rate

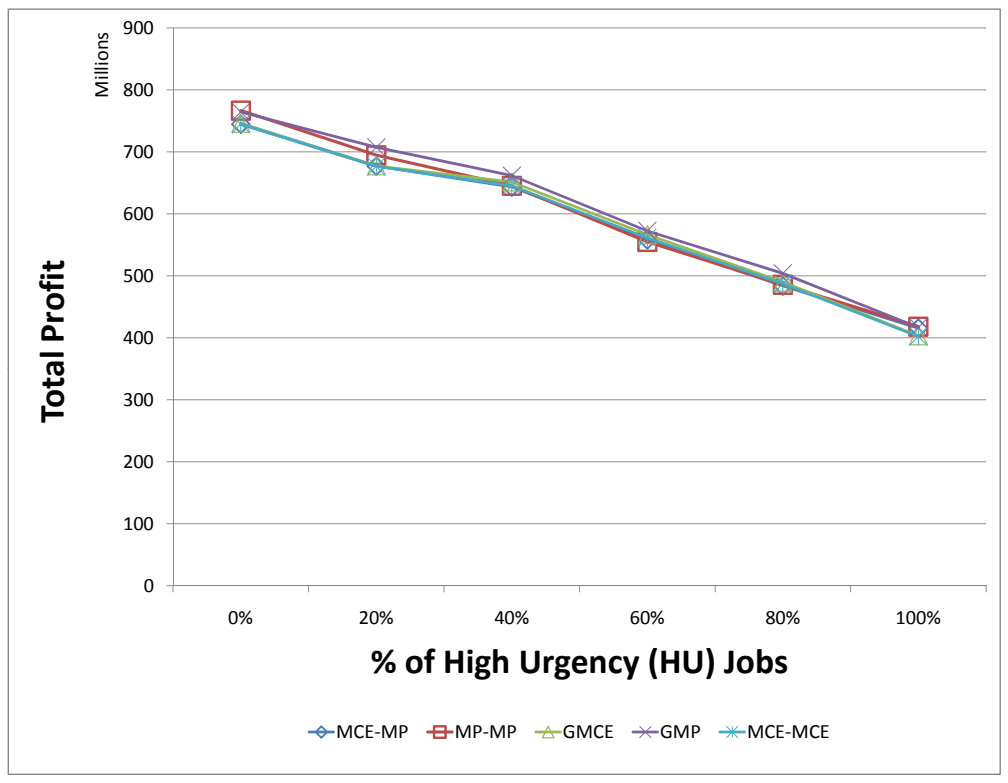

(c) Profit VS Urgency

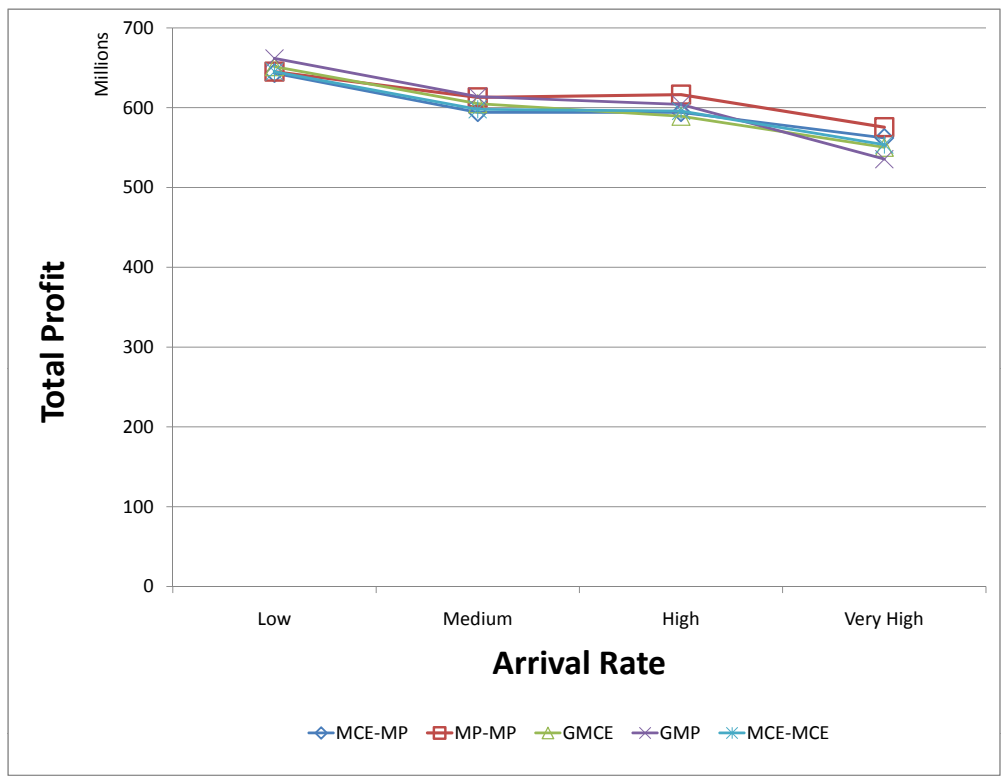

(d) Profit VS Arrival Rate

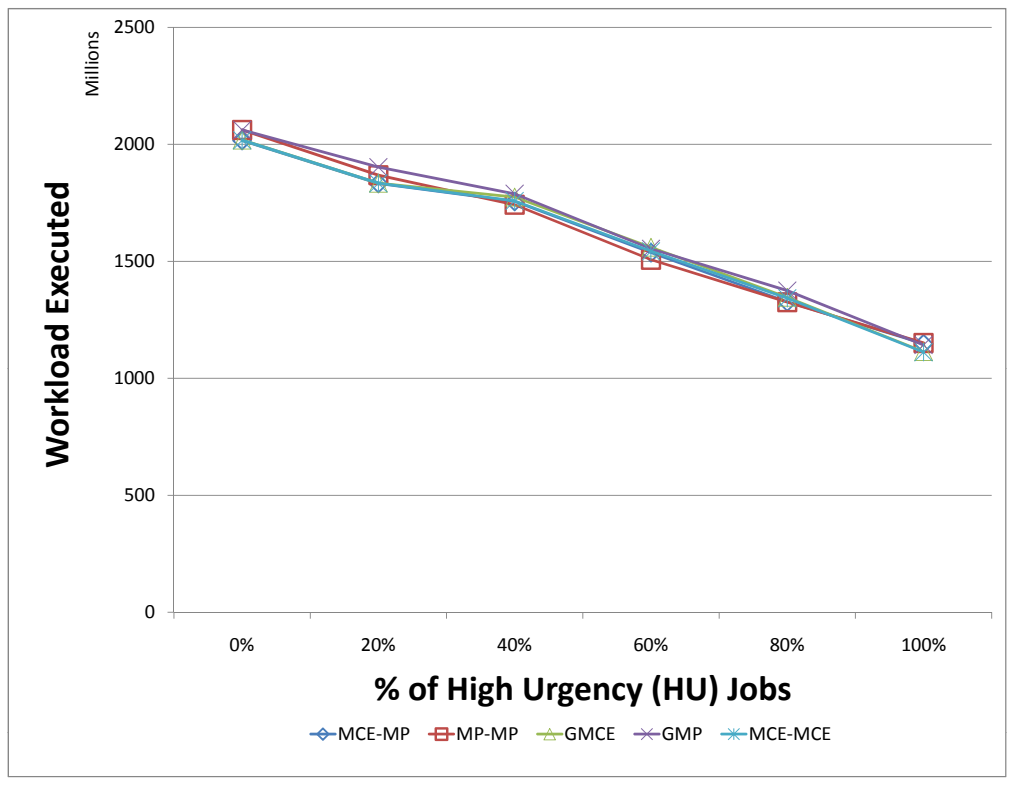

(e) Workload Executed VS Urgency

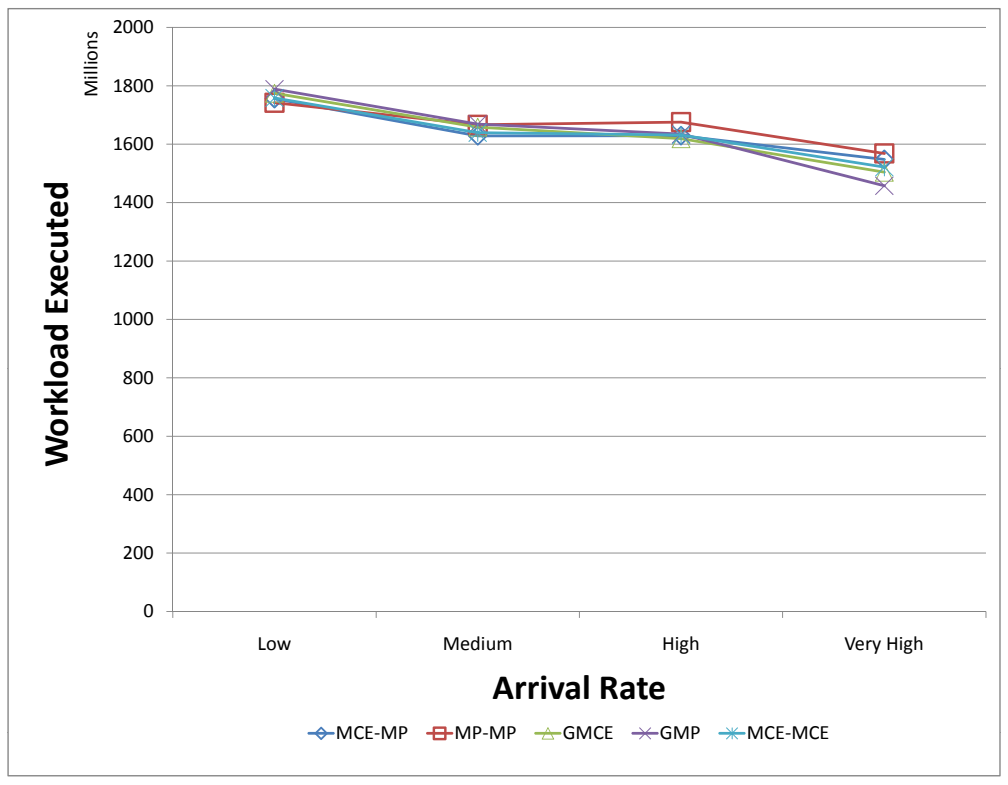

(f) Workload Executed VS Arrival Rate

Fig. 4. Effect of Urgency and Arrival Rate of Applications

collective gain/loss in carbon emission and profit across all data centers.

When the number of HU applications increases, the total profit of all policies (Figure 4(c)) decreases almost linearly by about 45% from 0% to 100% HU applications. Similarly, there is also a drop in total carbon emission (Figure 4(a)). This fall in total carbon emission and total profit is due to the lower acceptance of applications as observed in Figure 4(e). In Figure 4(a), the de-crease in total carbon emission for profit based policies (GMP and MP-MP) is much more than that of carbon emission based policies (MCE-MP, GMCE, and MCE-MCE). This is because carbon emission based policies schedule applications on more carbon emission efficient sites.

Likewise, the increase in arrival rate also affects the total carbon emission (Figure 4(b)) and total profit (Figure 4(d)). As more applications are submitted, less applications can be accepted (Figure 4(f)) since it is harder to satisfy their deadline requirement when workload is high.

### *6.2 Effect of Mapping Policy and DVS*

As discussed in Section 4, our meta-scheduling policies are designed to save energy at two phases, first at the mapping phase and then at the scheduling phase. Hence, in this section, we examine the effect of each phase on performance.

First, we examine the effect of the mapping phase by comparing meta-scheduling policies without the energy saving feature at the scheduling phase, i.e. DVS is not available at the local scheduler. Hence, we name the without-DVS version of the carbon emission based policy (GMCE) and profit based policy (GMP) as GMCE-WithoutDVS and GMP-WithoutDVS respectively. For various urgency of applications (Figure 5(a)), GMCE-WithoutDVS can prevent up to 10% $CO_2$ emission over GMP-WithoutDVS. For various arrival rate of applications (Figure 5(b)), GMCE-WithoutDVS can produce up to 23% less carbon emission than GMP-WithoutDVS. The corresponding difference in energy cost (Figure 5(c) and 5(d)) between them is very little (about 1–2%). This is because with the decrease in energy consumption due to the execution of HPC workload, both carbon emission and energy cost will automatically decrease. This trend still remains by comparing GMCE and GMP, both of which uses DVS at the scheduling phase.

In Figure 5(a), with the increase in the number of urgent applications, the difference in carbon emission between the carbon emission based policy (GMCE-WithoutDVS) and profit based policy (GMP-WithoutDVS) is decreasing. This is due to more applications being scheduled on less carbon efficient sites in order to avoid deadline misses. This is also the reason for all four policies to execute decreasing workload as the number of HU applications increases (Figure 5(e)).

Next, we examine the effect of the scheduling phase by comparing meta-scheduling policies with DVS (GMCE and GMP) and without DVS (GMCE-WithoutDVS and GMP-WithoutDVS). With DVS, the energy cost (Figure 5(c)) to execute HPC workload has been reduced on average by 33% when we com-

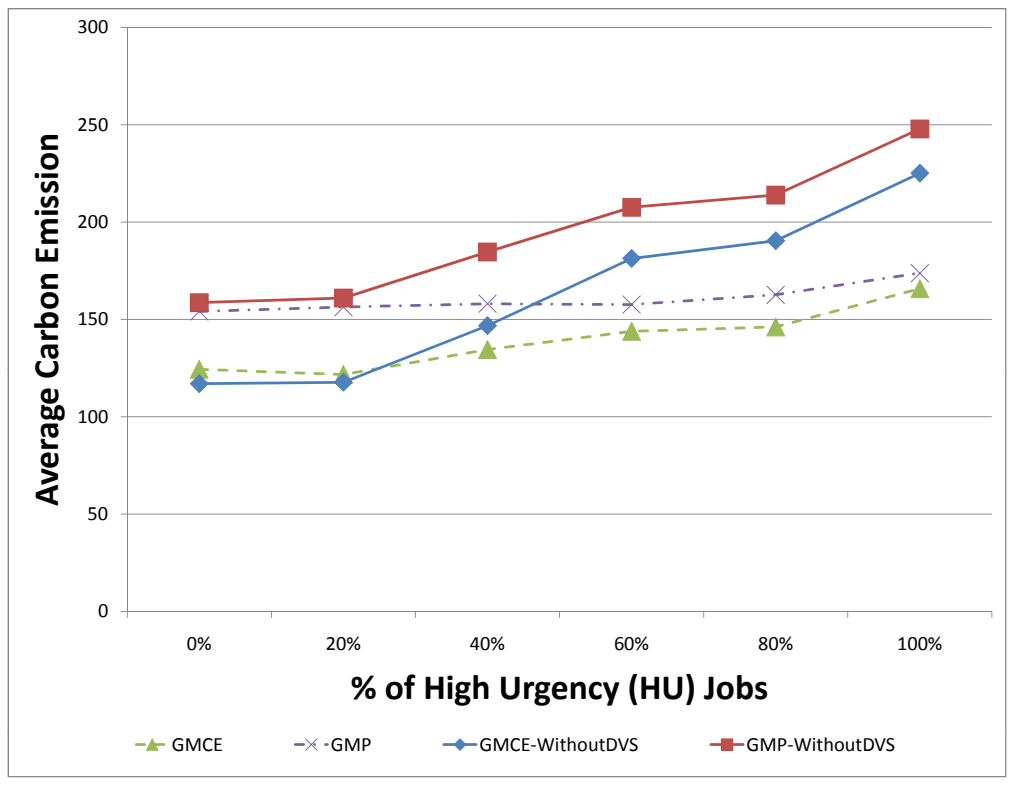

(a) Carbon Emission VS Urgency

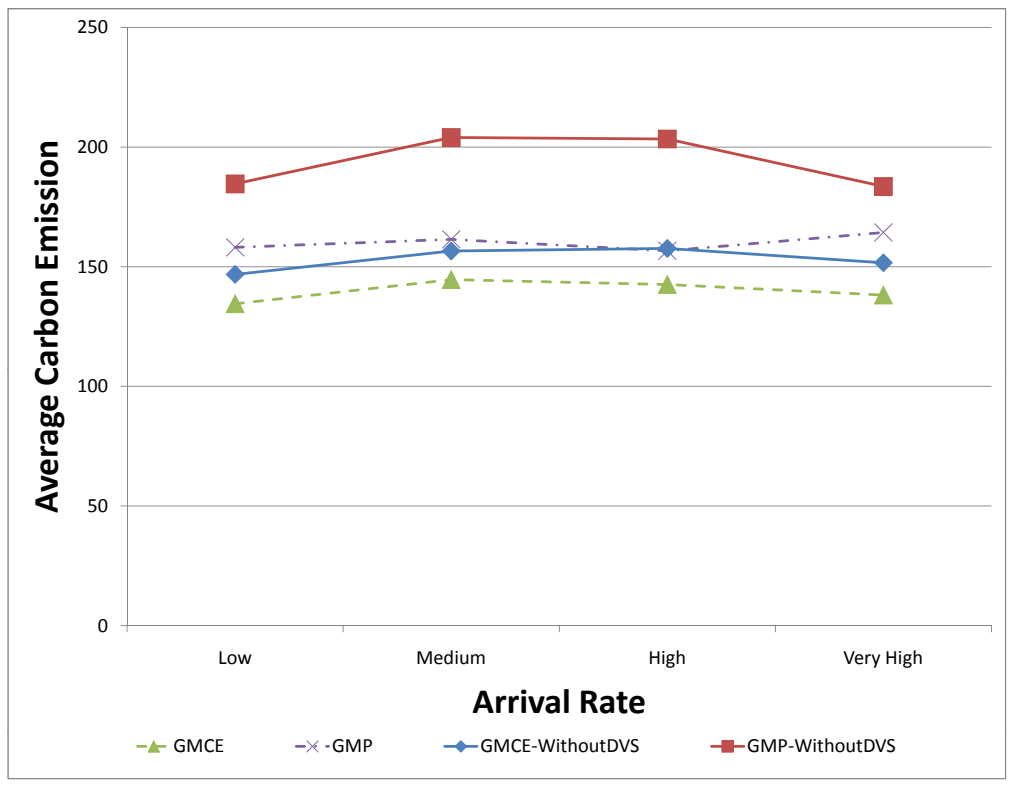

(b) Carbon Emission VS Arrival Rate

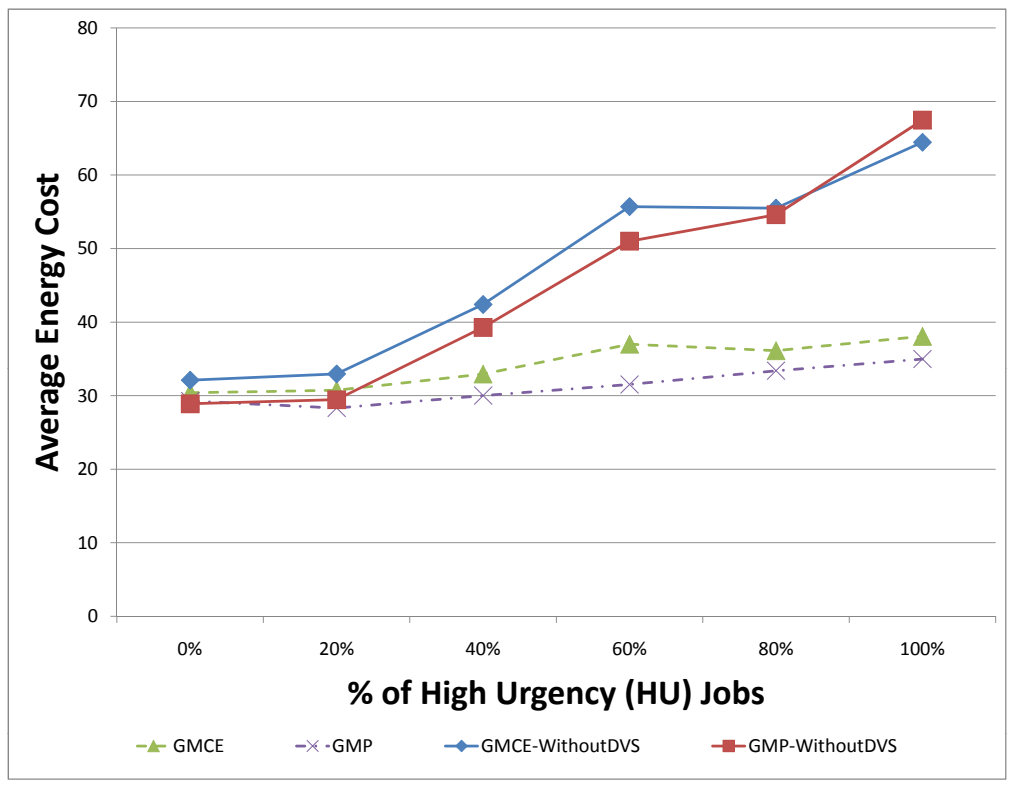

(c) Energy Cost VS Urgency

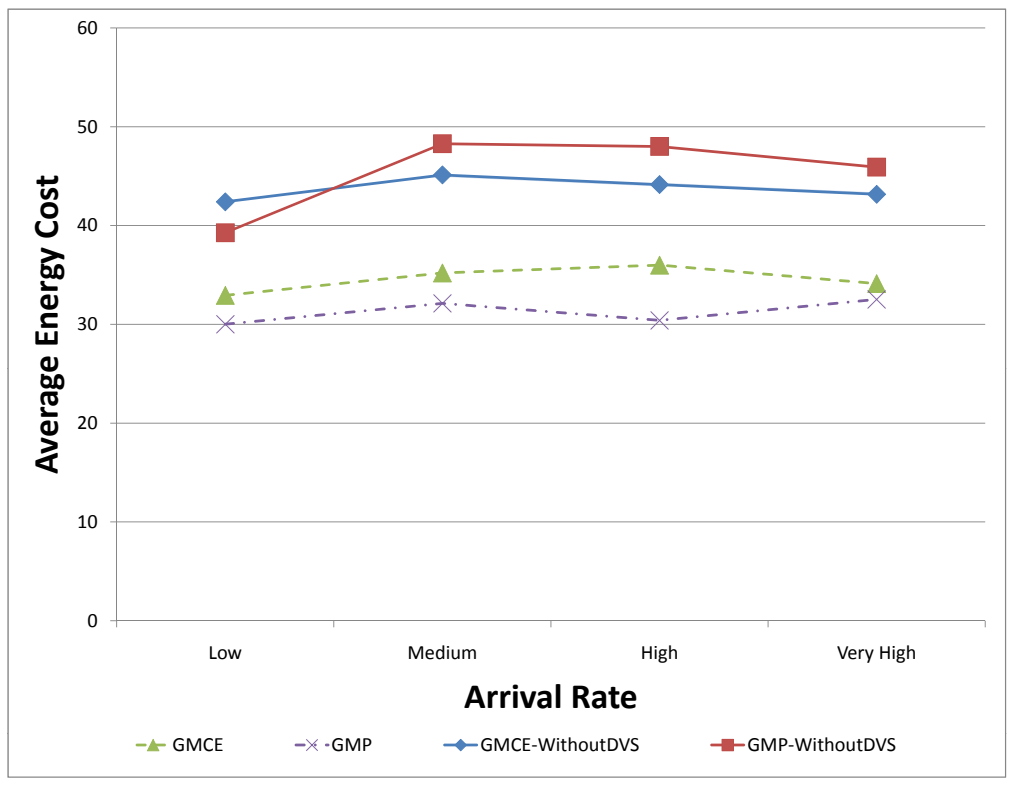

(d) Energy Cost VS Arrival Rate

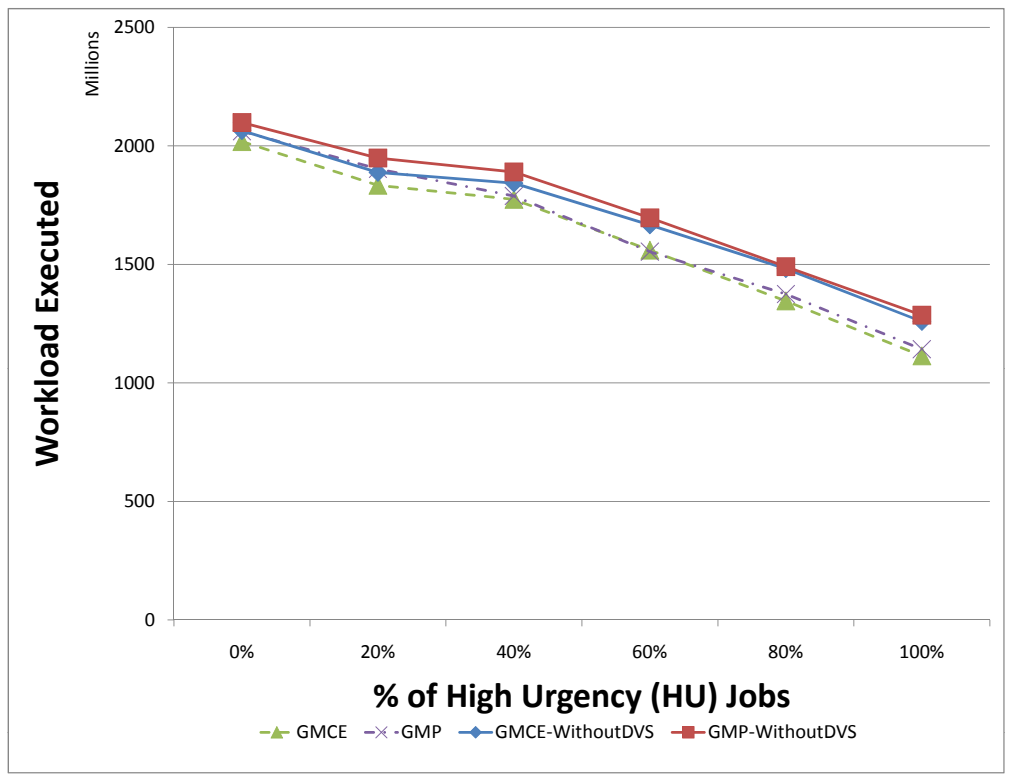

(e) Workload Executed VS Urgency

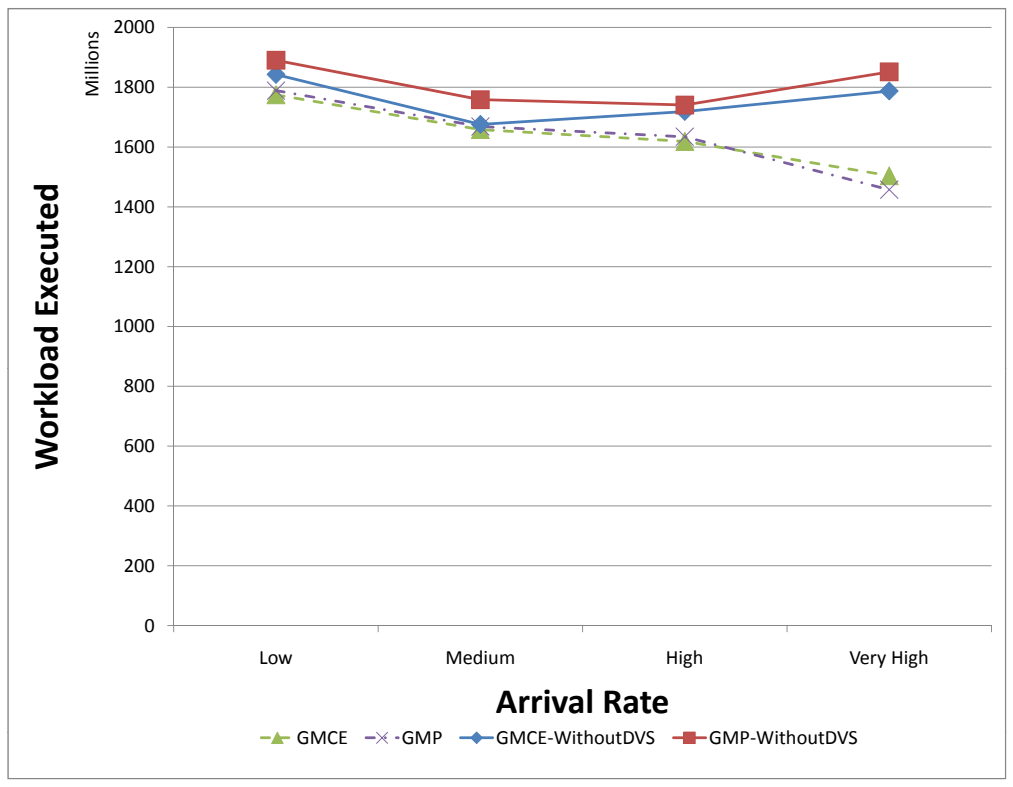

(f) Workload Executed VS Arrival Rate

Fig. 5. Effect of Mapping Policy and DVS

pare GMP with GMP-withoutDVS. With the increase in high urgency applications, the gap is increasing and we can get almost 50% decrease in energy cost as shown in Figure 5(c). With the increase in arrival rate, we get a consistent 25% gain in energy cost by using DVS (Figure 5(d)). The carbon emission is also reduced further on average by 13% with the increase in urgent applications as shown in Figure 5(a). With the increase in arrival rate, HPC workload executed is decreasing in the case of policies using DVS as can be observed from Figure 5(f). This is because the execution of jobs at lower CPU frequency results in more rejection of urgent jobs when the arrival rate is high. Thus, HPC workload executed in the case of policies without DVS is almost the same even when arrival rate is very high.

### *6.3 Impact of Carbon Emission Rate*

To examine the impact of carbon emission rate in different locations on our policies, we vary the carbon emission rate, while keeping all other factors such as electricity cost as the same. Using normal distribution with $mean = 0.2$, random values are generated for the following three classes of carbon emission rate across all Cloud sites as: A) Low variation (low) with $standard\ deviation = 0.05$, B) Medium variation (mid) with $standard\ deviation = 0.2$, and C) High variation (high) with $standard\ deviation = 0.4$. All experiments are conducted at medium job arrival rate with 40% of high urgency jobs.

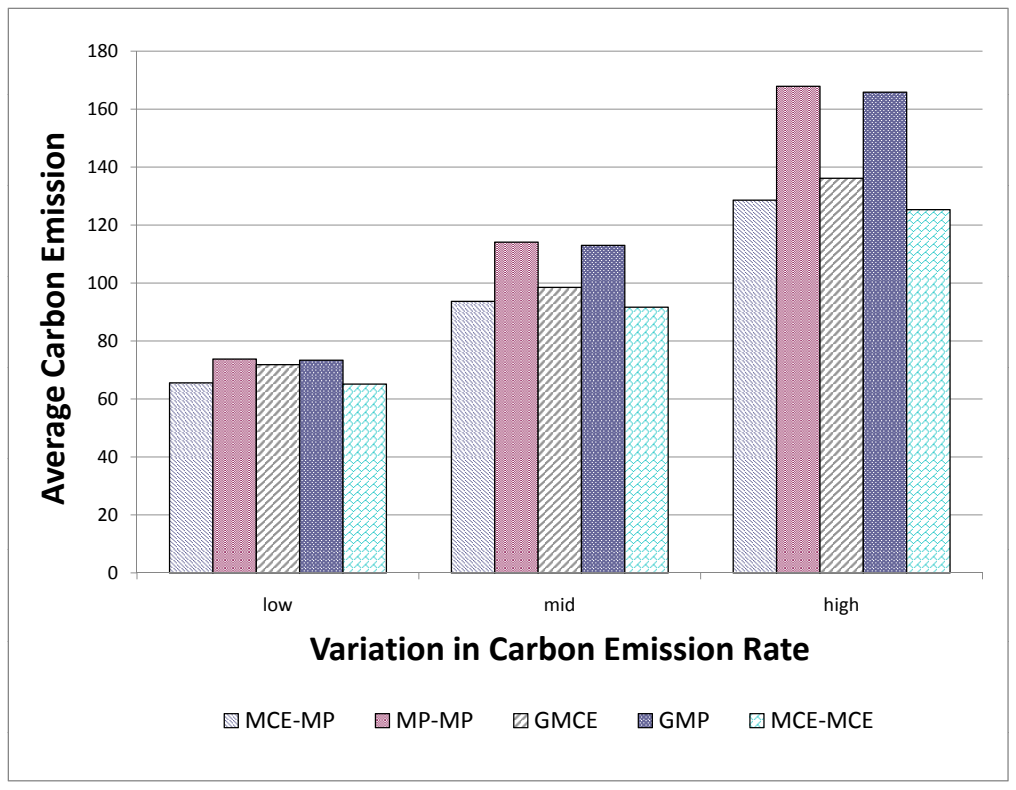

(a) Carbon Emission

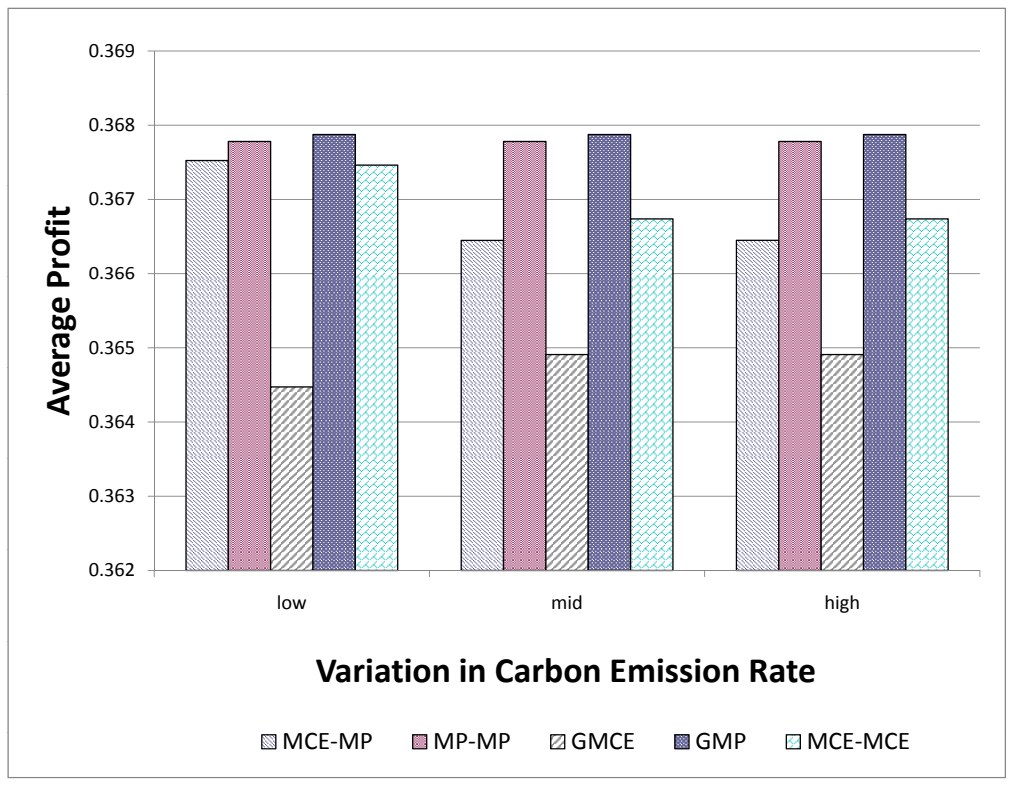

(b) Provider's Profit

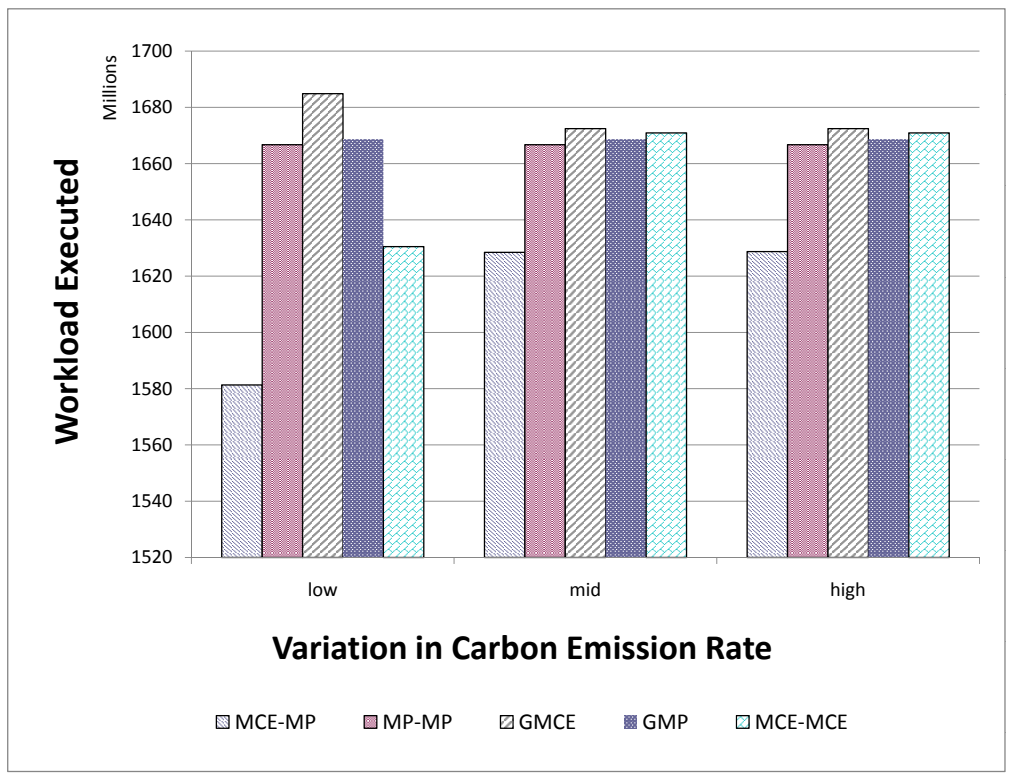

(c) Workload

Fig. 6. Impact of Carbon Emission Rate

The performance of all policies is similar for all three cases of carbon emission rate. For example, in Figure 6(a), the carbon emission of profit based policies (GMP and MP-MP) is always higher than carbon emission based policies (GMCE, MCE-MCE, and MCE-MP). Similarly, for profit (Figure 6(b)), all profit based policies perform better than all carbon emission based policies. For instance, in Figure 6(a), the difference in carbon emission of MCE-MCE and MP-MP is about 14% for low variation, which increases to 23% for high variation. On the other hand, in Figure 6(b), the corresponding decrease in profit is almost negligible which is less than 1%. Moreover, by comparing MCE-MCE and MP-MP in Figure 6(c), the amount of workload executed by MCE-MCE is slightly higher than MP-MP. Thus, for the case of high variation in carbon emission rate, Cloud providers can use carbon emission based policies such as MCE-MCE to considerably reduce carbon emission with almost negligible impact on their profit.

### *6.4 Impact of Electricity Cost*

To investigate the impact of electricity cost in different locations on our policies, we vary the electricity cost, while keeping all other factors such as carbon emission rate as the same. Using normal distribution with $mean = 0.1$, random values are generated for the following three classes of electricity cost/rate across all Cloud sites as: A) Low variation (low) with $standard\ deviation = 0.01$, B) Medium variation (mid) with $standard\ deviation = 0.02$, and C) High variation (high) with $standard\ deviation = 0.05$. All experiments are conducted at medium job arrival rate with 40% of high urgency jobs.

The variation in electricity cost/rate affects the performance of profit based policies (GMP and MP-MP) in terms of carbon emission (Figure 7(a)) and workload executed (Figure 7(c)), while carbon emission based policies (GMCE, MCE-MCE and MCE-MP) are not affected. But, the profit of all policies decrease more as the variation of electricity rate increases (Figure 7(b)) due to the subtraction of cost from profit. For high variation in electricity rate, there is not much difference (about 0.1%) in carbon emission between MP-MP and MCE-MCE (Figure 7(a)). Hence, Cloud providers can use MP-MP which gives slightly better average profit than carbon emission based policies (GMCE, MCE-MCE and MCE-MP). On the other hand, for cases when the variation in electricity rate is not high, providers can use carbon emission based policies such as MCE-MCE and MCE-MP to reduce about 1% of carbon emission by sacrificing less than 0.006% of profit.

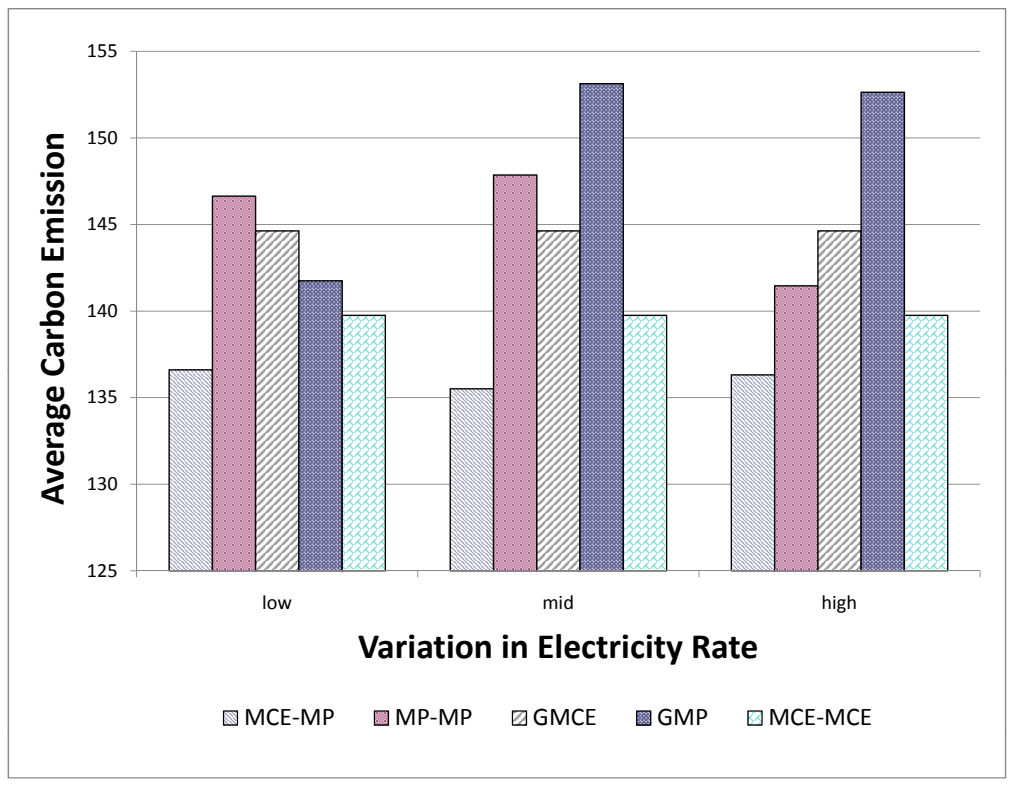

(a) Carbon Emission

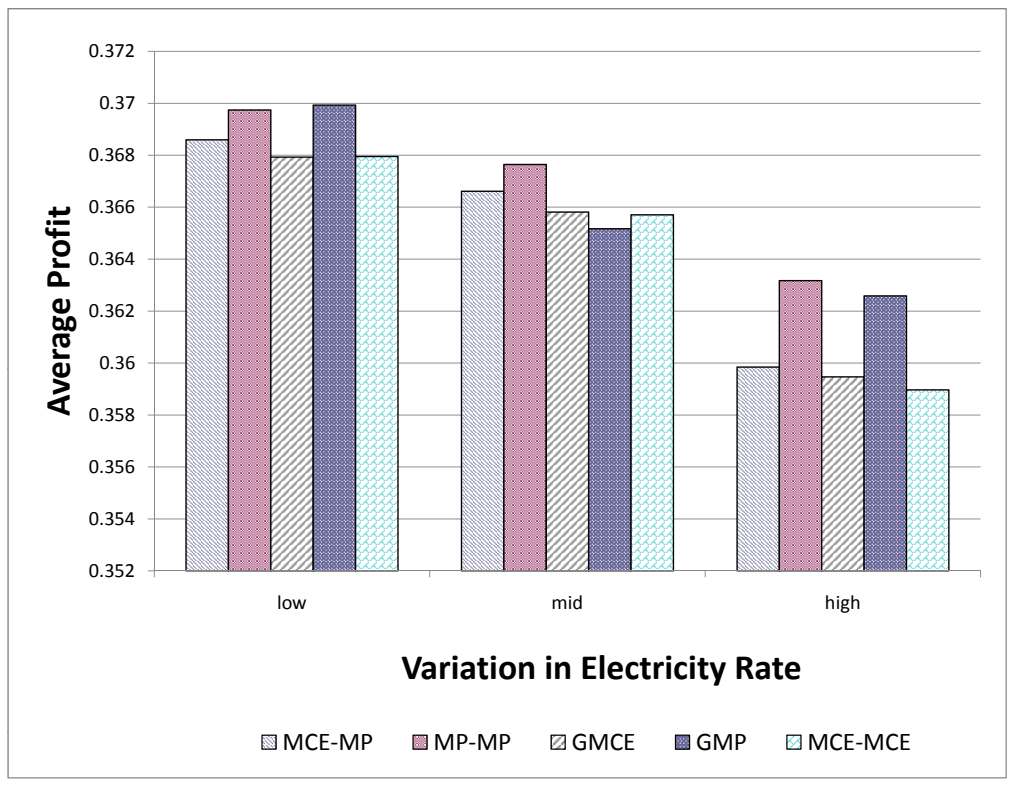

(b) Provider's Profit

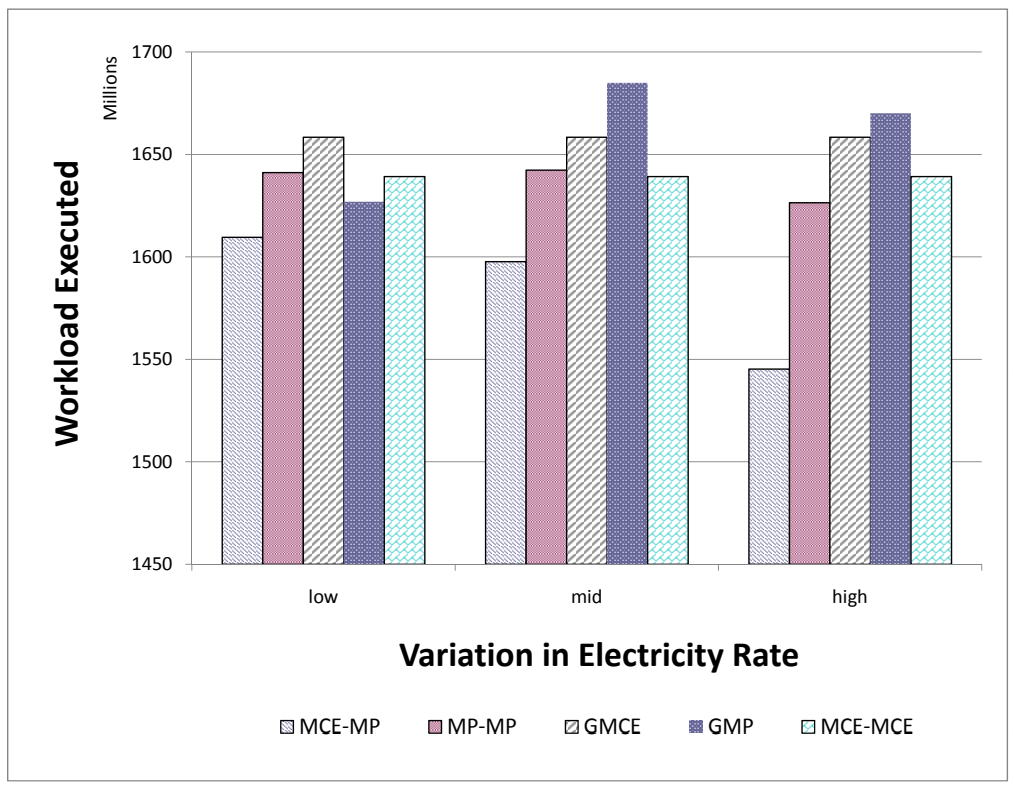

(c) Workload

Fig. 7. Impact of Electricity Cost

### *6.5 Impact of Data Center Efficiency*

To study the impact of data center efficiency in different locations on our policies, we vary the data center efficiency $= \frac{COP}{(COP+1)}$, while keeping all other factors such as carbon emission rate as the same. Using normal distribution with $mean = 0.4$, random values are generated for the following three classes of data center efficiency across all Cloud sites as: A) Low variation (low) with $standard\ deviation = 0.05$, B) Medium variation (mid) with $standard\ deviation = 0.12$, and C) High variation (high) with $standard\ deviation = 0.2$. All experiments are conducted at medium job arrival rate with 40% of high urgency jobs.

Figure 8(a) shows carbon emission based policies (GMCE, MCE-MCE and MCE-MP) achieve the lowest carbon emission with almost equal values. MCE-MCE performs better than MCE-MP by scheduling more HPC workload (Figure 8(c)) while achieving similar profit (Figure 8(b)). But when the variation in data center efficiency is high, GMCE can execute much higher workload (Figure 8(c)) than MCE-MCE and MCE-MP while achieving only slightly less profit than profit based policies (GMP and MP-MP) (Figure 8(b)). Thus, cloud providers can use GMCE to decrease the carbon emissions across their data centers without significant profit loss.

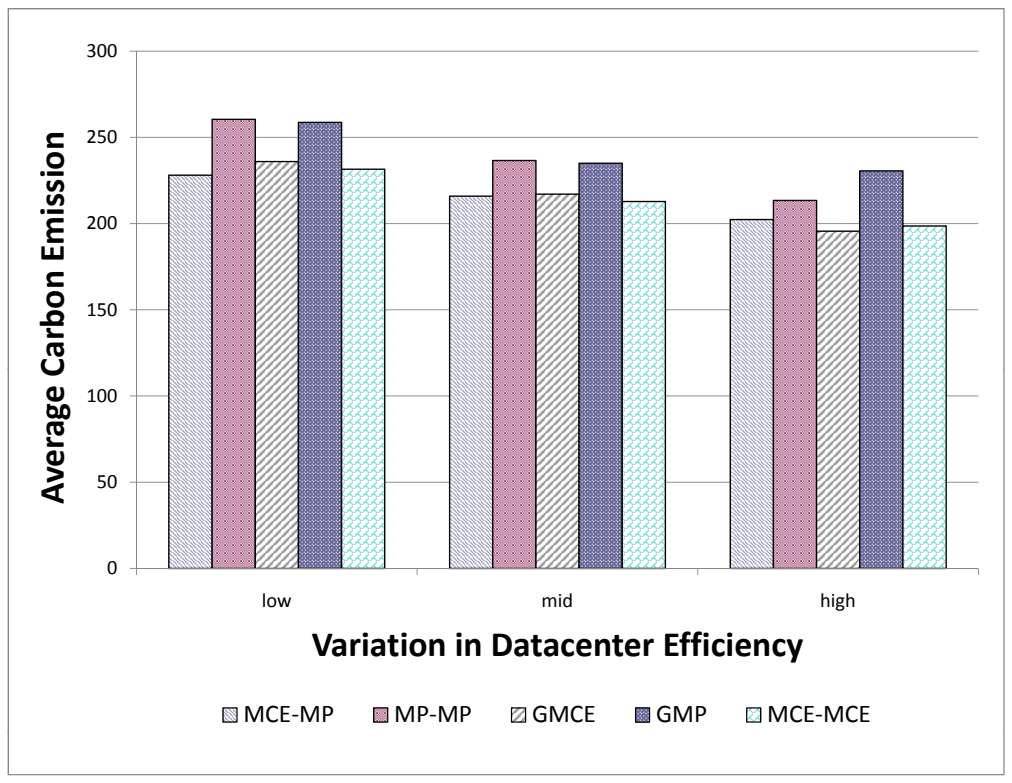

(a) Carbon Emission

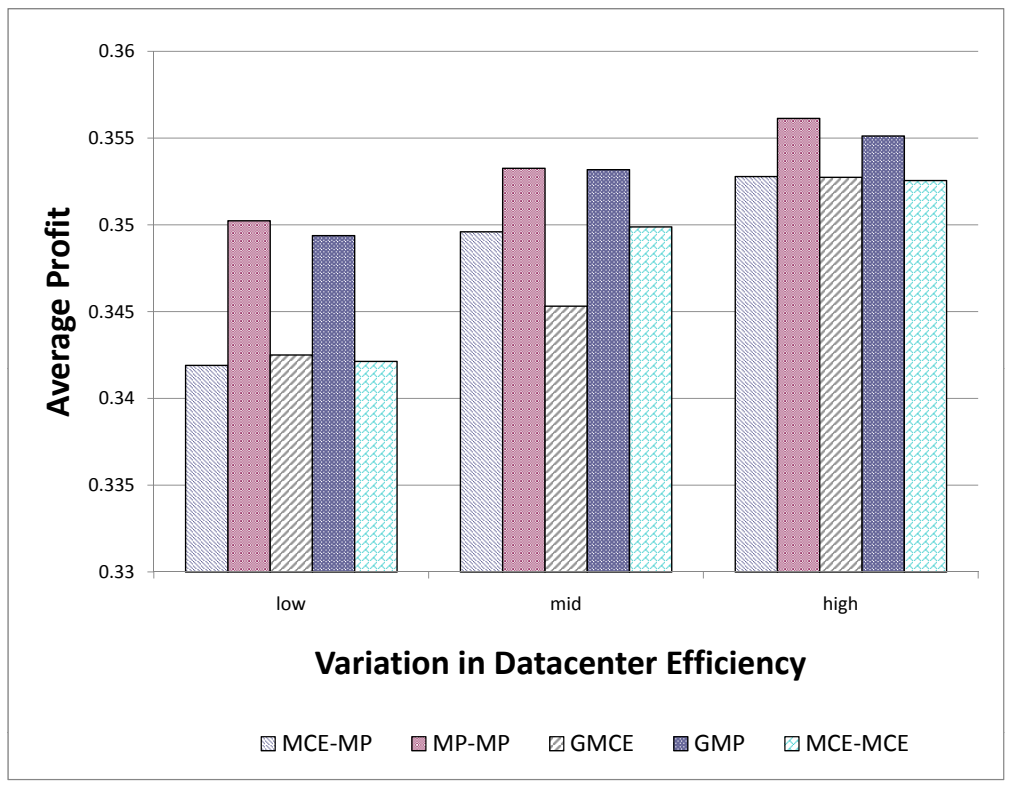

(b) Provider's Profit

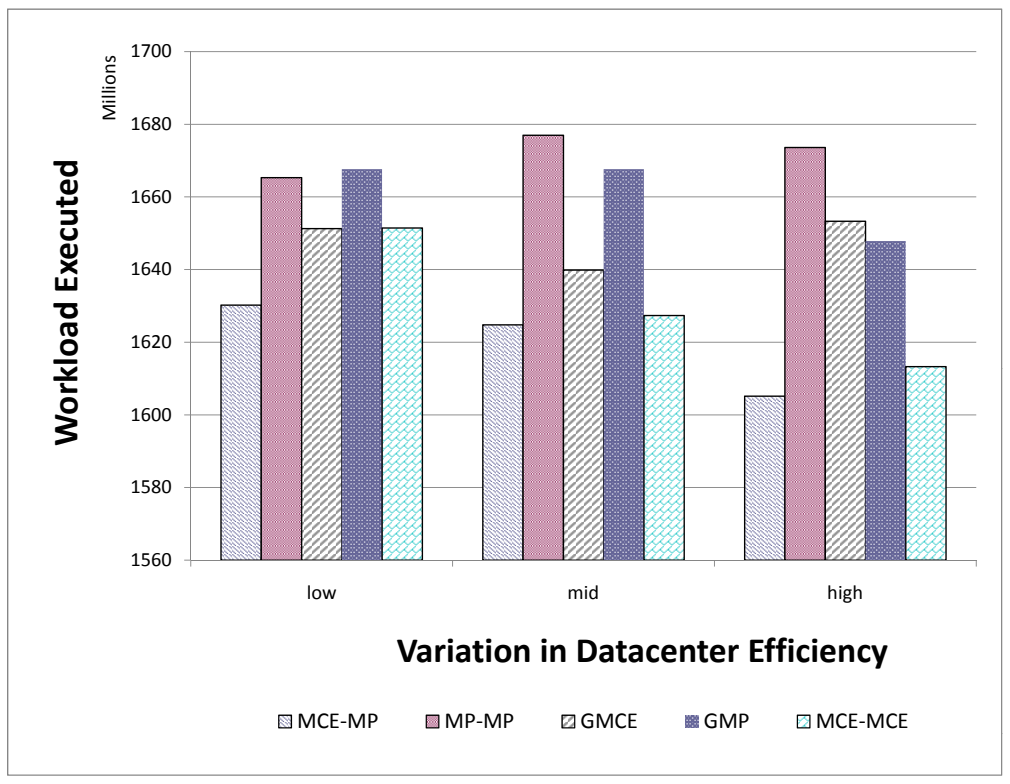

(c) Workload

Fig. 8. Impact of Data Center Efficiency

### *6.6 Comparison with Lower Bound and Upper Bound*

Figure 9 shows how different policies perform in comparison to the lower bound of average carbon emission and the upper bound of average profit. In Figure 9(a) and 9(b), with the increase in HU applications, the difference between the lower/upper bound and various policies is increasing. This is due to the increase in looseness of the bounds with the increase in HU applications. To avoid deadline misses with a higher number of HU applications, our proposed policies schedule more applications at higher CPU frequency which results in higher energy consumption. This in turn leads to an increase in the carbon emission and decrease in the profit. Whereas, for computing the lower/upper bounds, we only consider energy consumption at the optimal CPU frequency. Thus, the effect of urgency on the bounds is not as considerable as in our policies. This explains why our policies are closer to the bounds for a lower number of HU applications.

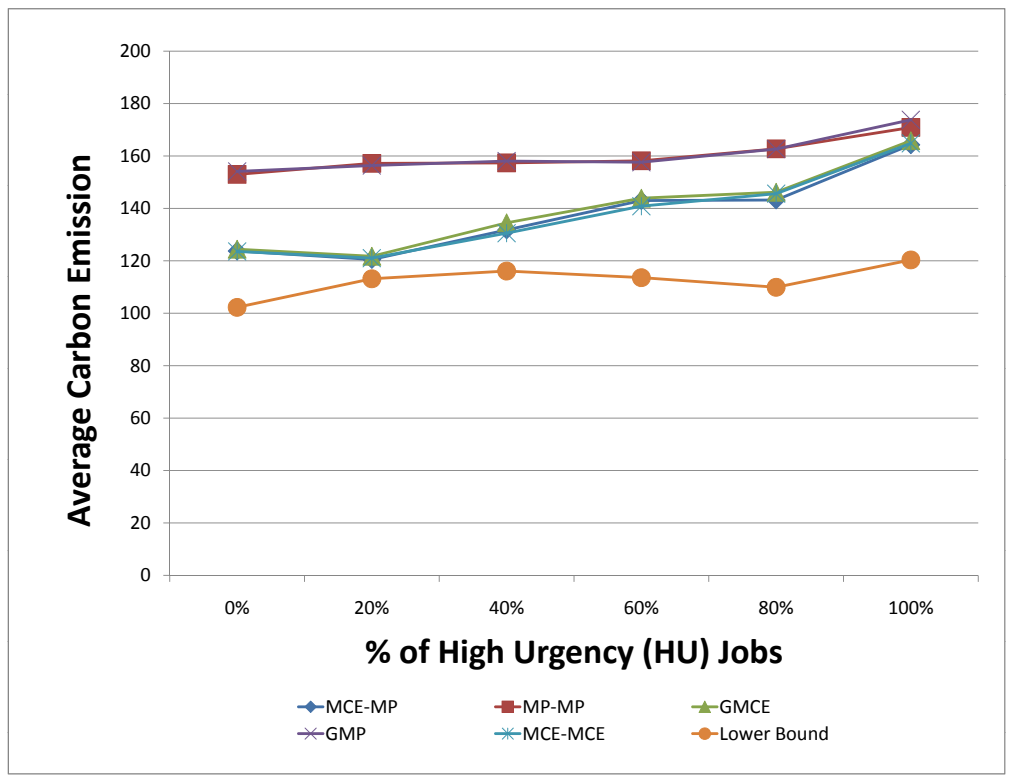

(a) Carbon Emission

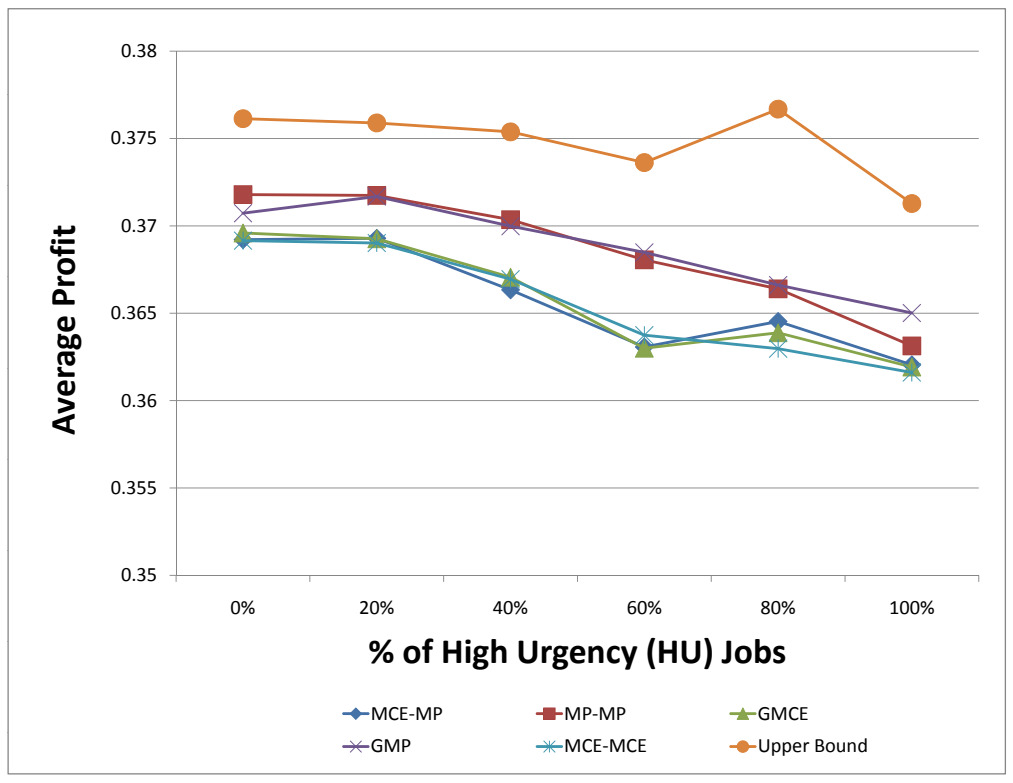

(b) Provider's Profit

Fig. 9. Comparison with Lower Bound and Upper Bound

In Figure 9(a), the difference in average carbon emission for carbon emission based policies (GMCE, MCE-MCE, and MCE-MP) and the lower bound is less than about 16% which becomes less than about 2% in the case of 20% HU applications. On the other hand, in Figure 9(b), the difference in average profit for profit based policies (GMP and MP-MP) and the upper bound is less than about 2% which becomes less than about 1% in the case of 40% of HU applications. Hence, in summary, our carbon emission based and profit based policies perform within about 16% and 2% of the optimal carbon emission and profit respectively.

### *6.7 Comparison of Our DVS and Previous DVS*

To correctly highlight the difference in DVS performance for the scheduling phase of the meta-scheduling policy, we need an independent policy (which is not linked to our proposed polices) for the mapping phase. Hence, we use EDF-EST, where the user jobs are ordered based on Earliest Deadline First (EDF), while the data centers are ordered based on Earliest Start Time (EST). Figure 10 shows that our proposed DVS (EDF-EST-withOurDVS) has not only outperformed previously proposed DVS [20][21](EDF-EST-withPrevDVS) by saving about 35% of energy, but also executed about 30% more workload. This is because most previously proposed DVS schemes focus on optimizing the power and thus try to run applications at the minimum CPU frequency $f_{min}$. As discussed in Section 4.2 and shown in Figure 3, it is clear that a job executed at $f_{min}$ may not lead to the least energy consumption due to the presence of local minima. Moreover, executing jobs at a lower frequency results in a lower acceptance of jobs since less CPUs are available.

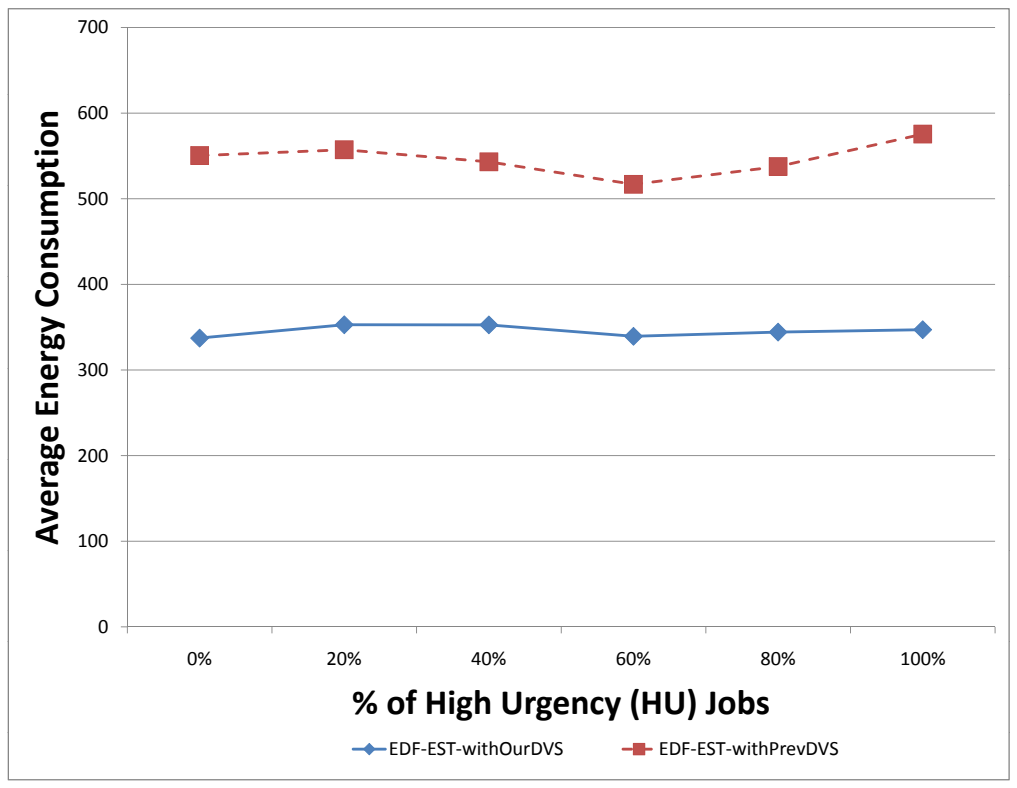

(a) Energy Consumption VS Urgency

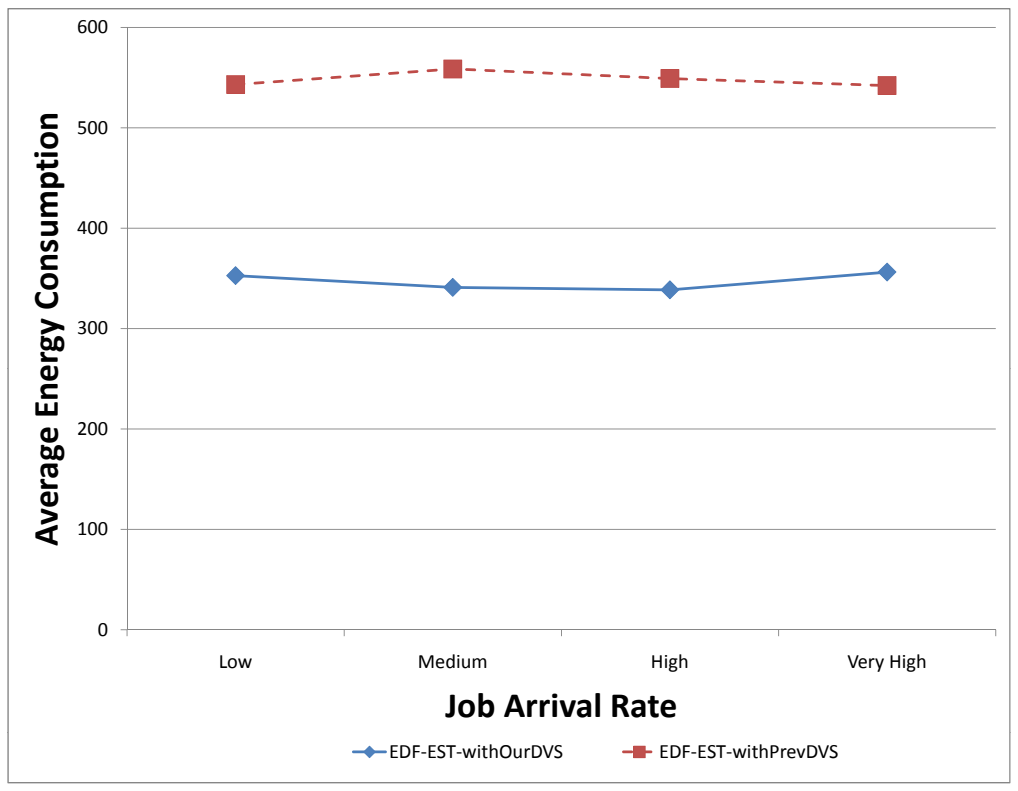

(b) Energy Consumption VS Arrival Rate

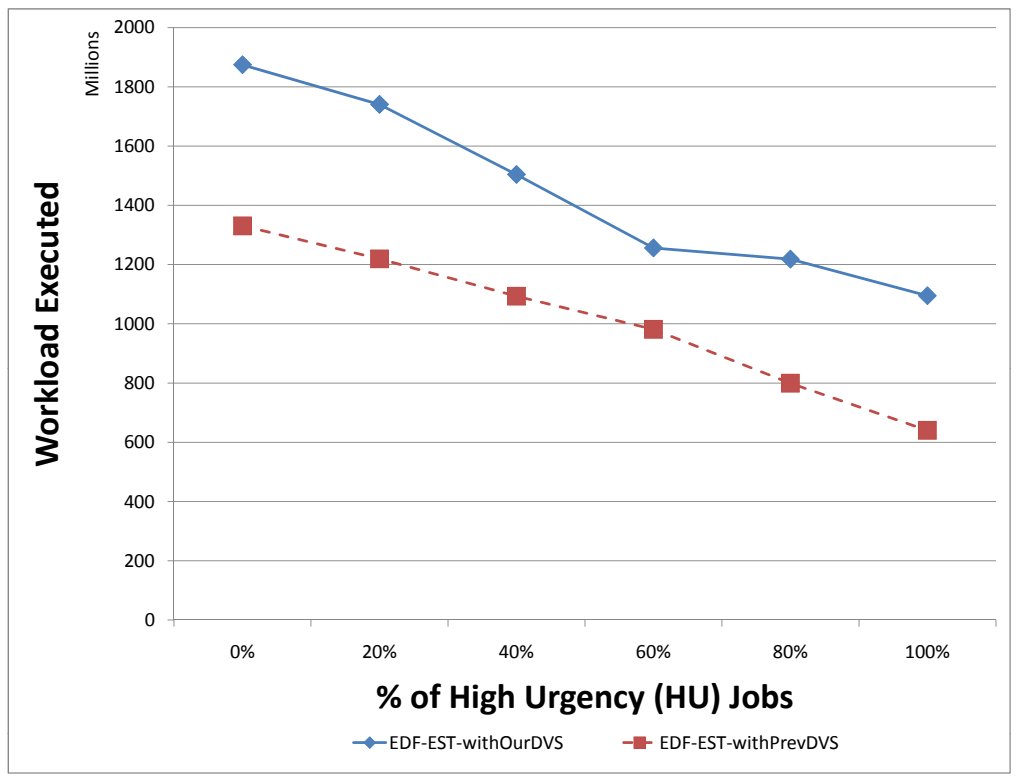

(c) Workload Executed VS Urgency

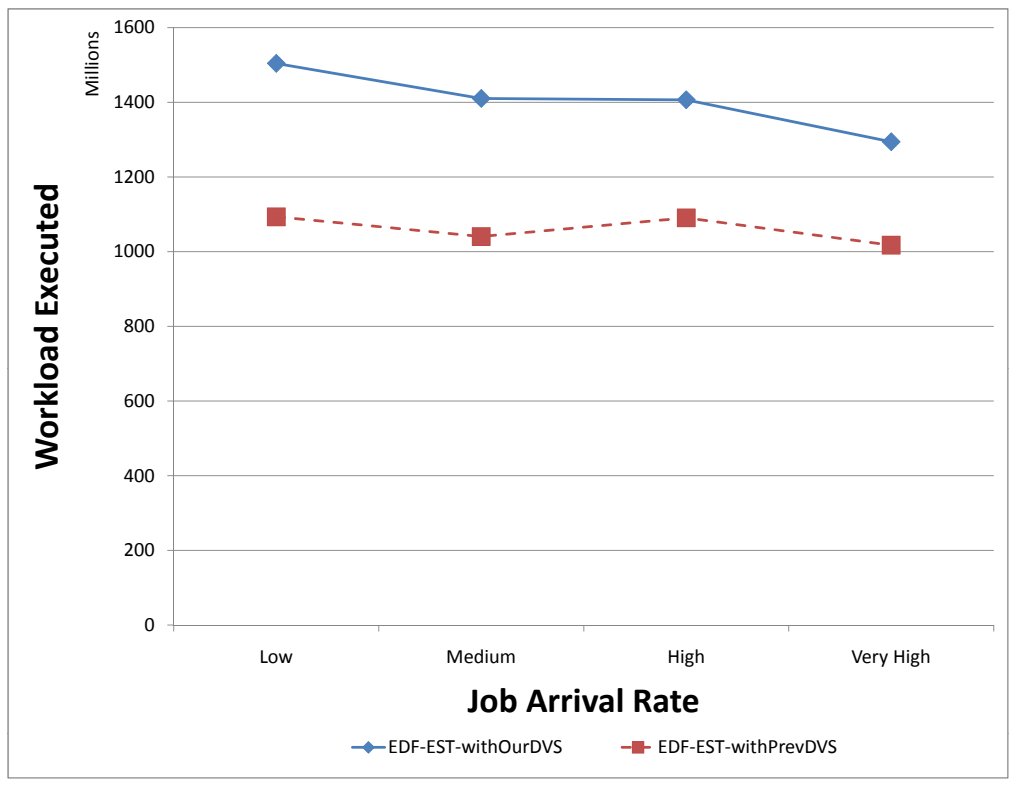

(d) Workload Executed VS Arrival Rate

Fig. 10. Effectiveness of Our Proposed DVS

## 7 Conclusions

The usage of energy has become a major concern since the price of electricity has increased dramatically. Especially, Cloud providers need a high amount of electricity to run and maintain their computer resources in order to provide the best service level for the customer. Although this importance has been emphasized in a lot of research literature, the combined approach of analyzing the profit and energy sustainability in the resource allocation process has not been taken into consideration.

The goal of this paper is to outline how managing resource allocation across multiple locations can have an impact on the energy cost of a provider. The overall meta-scheduling problem is described as an optimization problem with dual objective functions. Due to its NP hard characteristic, several heuristic policies are proposed and compared. The policies are compared with each other for different scenarios and also with the derived lower/upper bounds. In some cases, the policies performed very well with only almost 1% away from the upper bound of profit. By introducing Dynamic Voltage Scaling (DVS) and hence lowering the supply voltage of CPUs, the energy cost for executing HPC workloads can be reduced by 33% on average. Jobs will run on CPUs with a lower frequency than expected, but they still meet the deadlines. The limitation of carbon emission can be forced by governments to comply with certain threshold values [11]. In such cases, Cloud providers can focus on reducing carbon emission in addition to minimizing energy consumption. We identified that policies like MCE-MCE can help provider to reduce their emission while almost maintaining their profit. If the provider faces a volatile electricity rate, the MP-MP policy will lead to a better outcome. Depending on the environmental and economic constraints, Cloud providers can selectively choose different policies to efficiently allocate their resources to meet customers' requests. The characteristics and performance of each meta-scheduling policy are summarized in Table 3, where "low" and "high" represent the scenario for which the overall performance of the policy is given. For instance, GMCE performs the best when the variation in data center efficiency is high, while MCE-MP performs the best when the variation in energy cost is low or when there is a low number of HU applications.

Table 3
Summary of Heuristics with Comparison Results

| Meta-Scheduling Policy | Description | Time Complexity | Overall Performance | | | | |
|---|---|---|---|---|---|---|---|
| | | | HU Jobs | Arrival Rate | $CO_2$ Emission Rate | Data Center Efficiency | Energy Cost |
| GMCE | Greedy ($CO_2$ Emission) | $O(NJ)$ | Bad | Bad | Bad | Best (high) | Bad |
| MCE-MCE | Two-phase Greedy ($CO_2$ Emission) | $O(NJ^2)$ | Good (low) | Good (low) | Best (high) | Okay (low) | Good (low) |
| GMP | Greedy (Profit) | $O(NJ)$ | Okay (high) | Okay (high) | Bad (low) | Bad (high) | Bad |
| MP-MP | Two-phase Greedy (Profit) | $O(NJ^2)$ | Good (high) | Bad ($CO_2$ emission), Best (Profit) | Good (low) | Best (low) | Good (High) |
| MCE-MP | Two-phase Greedy ($CO_2$ Emission and Profit) | $O(NJ^2)$ | Best (low) | Good (high) | Okay | Okay | Best (low) |

In future, we will like to extend our model to consider the aspect of turning servers on and off, which can further reduce energy consumption. This requires a more technical analysis of the delay and power consumption for suspending servers, as well as the effect on the reliability of computing devices. We will also want to extend our policies for virtualized environments, where it can be easier to consolidate many applications on fewer physical servers.

## Acknowledgments

We would like to thank Marcos Dias de Assuncao for his constructive comments on this paper. This work is partially supported by research grants from the Australian Research Council (ARC) and Australian Department of Innovation, Industry, Science and Research (DIISR).